\documentclass[12pt]{article}

\usepackage{amsmath,amsfonts}
\usepackage{amssymb}
\usepackage{epsfig,psfrag,cite}
\usepackage[usenames,dvips]{color}
\usepackage{comment}
\usepackage{multirow,axodraw}

\usepackage{fullpage}

\makeatletter
\@addtoreset{equation}{section}
\makeatother

\newcommand{\be}{\begin{equation}}
\newcommand{\ee}{\end{equation}}
\newcommand{\bea}{\begin{eqnarray}}
\newcommand{\eea}{\end{eqnarray}}

\newcommand{\tr}{\operatorname{tr}}

\newcommand{\diag}{\operatorname{diag}}

\newcommand{\n}{\mathcal{N}}
\newcommand{\e}{\mathcal{E}}

\newcommand{\nn}{\nonumber}

\newcommand{\ceff}{c_\nu}

\begin{document}

\thispagestyle{empty}

\begin{center}
\hfill CPHT-RR 067-0812\\
\hfill UAB-FT-717

\begin{center}

\vspace{.5cm}

{\LARGE\sc  Neutrino Mixing from Wilson Lines in \\ \vspace{3mm}Warped Space
}

\end{center}

\vspace{1.cm}

{\bf Gero von Gersdorff$^{\;a}$, 
 Mariano Quir\'os$^{\,b}$, and Michael Wiechers$^{\;c}$}\\

\vspace{1.cm}
${}^a\!\!$ {\em {Centre de Physique Th\'eorique, \'Ecole Polytechnique and CNRS\\
F-91128 Palaiseau, France}}

\vspace{.1cm}

${}^b\!\!$ {\em {Instituci\'o Catalana de Recerca i Estudis  
Avan\c{c}ats (ICREA) and\\ Institut de F\'isica d'Altes Energies, Universitat Aut{\`o}noma de Barcelona\\
08193 Bellaterra, Barcelona, Spain}}

\vspace{.1cm}

${}^c\!\!$ {\em {Institut de F\'isica d'Altes Energies, Universitat Aut{\`o}noma de Barcelona\\
08193 Bellaterra, Barcelona, Spain}}

\end{center}

\vspace{0.8cm}

\centerline{\bf Abstract}
\vspace{2 mm}
\begin{quote}\small
We consider the generation of the hierarchical charged lepton spectrum and anarchic neutrino masses and mixing angles in warped extra dimensional models with Randall-Sundrum metric. We have classified all possible cases giving rise to realistic spectra for both Dirac and Majorana neutrinos. An anarchic neutrino spectrum requires a convenient bulk symmetry broken by boundary conditions on both UV and IR branes. We have in particular considered the case of Majorana neutrinos with a continuous bulk symmetry. To avoid unwanted massless extra gauge bosons the 4D group should be empty. If the 4D coset is not vanishing it can provide a Wilson Line description of the neutrino Majorana mass matrix. We have studied an example based on the bulk gauge group $U(3)_\ell\otimes U(3)_\n\otimes_i U(1)_{\e^i}$ with the Wilson Line in $SO(3)_\n$  satisfying all required conditions. A $\chi^2$-fit to experimental data exhibits 95\% CL region in the parameter space with no fine-tuning. As a consequence of the symmetries of the theory there is no tree-level induced lepton flavor violation and so one-loop processes are consistent with experimental data for KK-modes about a few TeV. The model is easily generalizable to models with IR deformed metrics with similar conclusions.  
\end{quote}

\vfill

\newpage

\section{Introduction}

A five-dimensional (5D) spacetime with a single warped extra dimension (WED)~\cite{Randall:1999ee}, and two branes localized respectively in the ultra-violet (UV) and infra-red (IR) regions, is an extensively explored alternative to supersymmetry as a possible solution to the Standard Model (SM) gauge hierarchy problem, provided that the Higgs field is sufficiently localized towards the IR boundary. Moreover this theory has, by means of the AdS/CFT correspondence, a description in terms of a dual four-dimensional (4D) strongly coupled theory by which 5D fields localized towards the IR boundary correspond to composite states in the dual theory while fields localized towards the UV brane correspond to elementary ones. While both, supersymmetric and WED theories, can accommodate a Higgs boson with a mass around 125 GeV, as that found by the recent LHC Higgs searches, its couplings to the different SM fields depend to a large extent on the different models. The comparison with experimental results must however wait until more accurate data on the different Higgs decay channels become available. 

On the other hand WED theories can successfully accommodate a solution to the flavor problem if all SM fields propagate in the bulk of the fifth dimension with light (heavy) fermion profiles leaning towards the UV (IR) boundary which are then mostly elementary (composite) states. This program has been very successfully applied to the quark sector in WED theories both for an AdS metric~\cite{Gherghetta:2000qt,Huber:2000ie}, as in the original Randall-Sundrum (RS) model, and for asymptotically AdS (or IR deformed) metrics~\cite{Carmona:2011ib,Archer:2011bk,Cabrer:2011qb,Archer:2012qa}. The reason of this success is that both the quark spectrum (in the up and down sectors) and the CKM mixing angles are hierarchical, a situation which can be readily described for order unity 5D Yukawa couplings if the various quark flavors are differently localized along the fifth dimension. In fact quark localization, and thus its 4D Yukawa coupling (determined by its overlapping with the Higgs profile), is controlled by their 5D Dirac mass $c$ such that
for different values of $c$ (mainly in the left-handed doublet and right-handed singlet up sectors) we obtain hierarchically different values of the corresponding quark masses and small mixing angles. Moreover due to their different localizations the quark flavors couple differently to KK modes of gauge bosons (e.g.~gluinos) thus generating dimension-six flavor changing neutral current (FCNC) operators. While these flavor violations are suppressed by the so-called RS-GIM mechanism~\cite{Agashe:2004ay} the KK masses have to be heavier than $\sim$ 20 TeV to suppress FCNC processes in the RS theory~\cite{Csaki:2008zd,Cabrer:2011qb} while this bound can be lowered to 4-5 TeV in the case of IR deformed metrics~\cite{Archer:2011bk,Cabrer:2011qb}. 

While most of the flavor literature in WED has been devoted to the quark sector, the lepton sector has also been extensively explored~\cite{Grossman:1999ra,Kitano:2000wr,Huber:2002gp,Huber:2003tu,Huber:2003sf,Gherghetta:2003he,Chang:2005ag,Moreau:2005kz,Agashe:2006iy,Chen:2008qg,Perez:2008ee,Csaki:2008qq,Agashe:2008fe,Agashe:2009tu,delAguila:2010vg,Atkins:2010cc,Kadosh:2010rm,Hagedorn:2011un,Csaki:2010aj,Watanabe:2010cy}. In fact the lepton sector is qualitatively different from the quark sector as the charged lepton spectrum is hierarchical while the neutrino spectrum and PMNS mixing angles are mostly anarchical~\cite{Hall:1999sn}. So while different  lepton localization is still  useful  to describe the charged lepton spectrum it is not that much in order to describe the spectrum and mixing angles in the neutrino sector. Another characteristic feature of the lepton sector is the neutrino nature, i.e.~Dirac versus Majorana, which is obviously related to the possibility of lepton number violation in the bulk and/or the branes of the 5D theory. In fact the presence (or absence) of lepton number violation is mostly related to the existence of (possibly gauged because of the AdS/CFT correspondence) symmetries in the bulk and the boundaries and will be the main feature of WED theories aiming to describe the theory of leptons. Moreover the different lepton flavor localization generates, by the tree-level exchange of KK modes of electroweak gauge bosons, lepton flavor violation (LFV) as dimension-six operators corresponding to processes as $\mu\to 3e$ and $\mu-e$ conversion, or in loop-diagrams from exchanged KK modes of charged leptons, neutrinos and the Higgs boson, in processes as $\mu\to e\gamma$. Actually there is a tension  between LFV in tree-level processes, which puts lower bounds on the Yukawa elements (the larger the Yukawas the more localized towards the UV brane the leptons and the more effective the RS-GIM mechanism), and the one-loop processes which put upper bounds on the Yukawa entries (the smaller the Yukawas the smaller the one-loop results as they are proportional to chirality-flipping mass insertions)~\cite{Agashe:2006iy}.

In this paper we will construct a WED theory of leptons where the spectrum of charged leptons is hierarchical while the spectrum and mixing angles in the neutrino sector are anarchical. We will separately review the cases where 4D neutrinos are Dirac and Majorana fermions and we will consider, and independently classify, all possibilities with special emphasis in those cases where fermion localization can successfully lead to realistic neutrino spectra. The former case (Dirac neutrinos) is similar to the description of the quark sector while the case of Majorana neutrinos is done by computing the coefficient of the dimension-five Weinberg operator obtained by integrating out the right-handed neutrinos in the 5D diagram $H\ell_i\to H\ell_j$ where right-handed neutrinos are exchanged in the s-channel. This integration is the 5D equivalent of the seesaw mechanism in 5D theories. The general result (both for Dirac and Majorana neutrinos) is that if the WED theory solves the $M_P$/TeV hierarchy, a realistic anarchic neutrino spectrum strongly motivates the presence of a 5D symmetry which ensures that $c_\ell^i\equiv c_\ell$ and $c_\n^j\equiv c_\n$ for $\forall i,j$, where $c_\ell^i$ and $c_\n^j$ are the 5D Dirac masses for lepton doublets $\ell_i$ and right-handed neutrinos $\n_j$. Moreover we have found for the Majorana case that if lepton number is broken in the bulk of the fifth dimension (whether or not it is broken in the UV and/or IR brane) it is not possible to accommodate the spectrum of light neutrinos and charged leptons simultaneously. Otherwise if lepton number is conserved in the bulk a realistic neutrino sector can be obtained if the theory is such that lepton number is violated in the UV brane, as originally proposed in \cite{Huber:2003sf}.

The remainder of this paper is devoted to proposing a symmetry which ensures the degenerate spectrum and large mixing angles for the neutrino sector. We will work out the case of a continuous gauge symmetry $G$ in the bulk broken to the subgroup $H_0$ ($H_1$) on the UV (IR) brane by boundary conditions. This symmetry guarantees the required structure on the 5D Dirac masses as well as the required Yukawa matrices for the different sectors, while in the 4D theory no massless gauge zero mode remains. Furthermore for the case of Majorana neutrinos we have proposed a Wilson Line (WL) model as the origin of the $U_{PMNS}$ matrix which is generated by a background value along the coset $K=K_0\cap K_1$ where $K_i$ is the coset $G/H_i$. Using $\mathcal O(1)$ values of the 5D Yukawas we have made a fit to the parameters which control the WL 
and shown the 95\% CL regions in the WL parameters which exhibit no fine-tuning in the determination of the anarchic mixing angles and neutrino masses. Our model can incorporate both a regular hierarchy and a inverted hierarchy of neutrino masses. Finally we have considered in our model the issue of LFV. As a matter of fact, due to the structure of the Yukawas and 5D masses imposed by the symmetry, there is no contribution to tree-level processes as $\mu\to 3e$ and therefore there is no tension between tree-level and loop induced processes as $\mu\to e\gamma$. In fact the latter can be comfortably below experimental bounds for KK masses around 2-3 TeV and the IR Yukawa couplings of $\mathcal O(1)$.

The plan of this article is as follows. In Sec.~\ref{Dirac} we have considered the case of conserved lepton number, i.e.~Dirac neutrinos. In Sec.~\ref{Majorana} the case of Majorana neutrinos is worked out in detail. The 5D propagator for right-handed neutrinos at zero-momentum is computed and a general expression for the coefficient of the 4D Weinberg operator explicitly given in the presence of lepton number violation in the bulk and in the branes for an arbitrary number of right handed neutrinos. In Sec.~\ref{cMnonzero} this general result where lepton number is violated in the bulk (and the UV and IR branes) of the fifth dimension is applied for simplicity to the case of only one generation of 5D right-handed neutrinos as (unlike in the 4D theories) the three left-handed neutrinos can receive Majorana masses by the 5D seesaw mechanism. However, as we have previously noticed, this case is unrealistic if one wants to describe the neutrino spectrum without introducing new scales, as the neutrino masses turn out to be too small. In Sec.~\ref{cMzero} we have considered the case of an arbitrary number of right-handed neutrinos where lepton number is conserved in the bulk, and violated on the IR and UV branes. In this case the leading contribution from the previous case vanishes and the subleading correction can provide realistic values of the neutrino spectrum without any fine-tuning. As noticed above, describing the degenerate spectrum and large mixings of the neutrino data requires a symmetry. In Sec.~\ref{flavor} we have worked out a particular gauge symmetry in the bulk, consistent with an anarchic structure for the neutrino spectrum and mixing angles. The bulk gauge symmetry is broken on the branes by boundary conditions with no zero mode in the 4D theory and such that the matrix diagonalizing the neutrino mass matrix, $U_{PMNS}$, is determined by a Wilson Line along the coset of the broken symmetry. In Sec.~\ref{LFV} we have examined LFV in the present theory and shown that no tree-level LFV processes are generated as a consequence of the underlying symmetry while loop level induced processes, such as $\mu\to e\gamma$, are  shown to give mild constraints on KK masses and the 5D Yukawa couplings. Finally Sec.~\ref{conclusion} is devoted to our conclusions. We present in App.~\ref{details} details on the calculation of the general 5D propagator of a right-handed neutrino in the presence of lepton number violation in the bulk and the IR and UV branes. In App.~\ref{moredetails} we provide details of the 5D right-handed neutrino propagator for an arbitrary number of right-handed neutrinos,  for lepton number conserved in the bulk and violated in the UV and IR branes.

\section{Dirac neutrinos}
\label{Dirac}

Let us denote the background metric by
\be
ds^2=e^{-2A}\eta_{\mu\nu}dx^\mu dx^\nu+dy^2\,,\qquad \eta_{\mu\nu}=\diag(-1,+1,+1,+1)\,.
\ee
We will consider leptons as 5D Dirac fermions propagating in the bulk, with
\be
\ell_i(x,y)
, \quad 
\e_i(x,y)
, \quad 
\n_i(x,y)\ 
\label{leptones}
\ee
denoting lepton doublets, singlets and right-handed (RH) neutrinos respectively. Here we use $i$ to label the three generations of $\ell_i$ and $\e_i$, and the (a priori arbitrary) number of copies of $\n_i$.
The kinetic Lagrangian for an arbitrary metric~\cite{Cabrer:2011qb} reads~\footnote{We are using a notation where the left-handed components of the 4D $SU(2)_L$ doublets are described by $\ell_L$ while the right-handed components of the 4D neutrino and lepton singlets are described by $\n_R$ and $\e_R$, respectively.}
\be
\mathcal L_{\rm kin}= \sum_{\psi=\ell,\n,\e}\int dy\,\left[e^{-3A}\left(i\bar\psi_L\, /\hspace{-.22cm}\partial\,\psi_L
+i\bar\psi_R\, /\hspace{-.22cm}\partial\,\psi_R\right)
+
e^{-4A}\,\left(-
\bar\psi'_R\psi_L+2A'\,\bar\psi_L\psi_R+{\rm h.c.}\right)\right]
\label{A1}
\ee
and parametrize the 5D mass Lagrangian as
\be
\mathcal L_{\rm mass}=-\sum_{\psi=\ell,\n,\e}\int dy\, e^{-4A} c_\psi \,M(y)\,\bar\psi_R\psi_L+h.c.
\ee
with constants
$
c_\psi=(-c_\ell,c_\n,c_\e)
$
and $M(y)$ a function with the dimension of mass. Although everything can be easily worked out for arbitrary metric $A(y)$ and mass profile $M(y)$, we will for the sake of simplicity 
specialize to the AdS case $A(y)=k\,y$ and constant mass profile $M(y)=k$.
With appropriate boundary conditions there  are then zero modes with profiles
\be
f_{\psi_\chi}^{(0)}(y)=\frac{e^{(2-c_\psi)ky}}{N_\psi^{1/2}}\,,
\qquad
N_\psi=\frac{e^{(1-2c_\psi)ky_1}-1}{(1-2c_\psi)k}
\label{fermion}
\ee
%
for  $\psi_\chi=\ell_L,\ \e_R,\ \n_R$  while the wave functions for the opposite chiralities 
($\ell_R,\ \e_L,\ \n_L$) vanish identically. 

We will now consider a 5D Yukawa interaction between a bulk Higgs field $H(x,y)$ and the leptons in Eq.~(\ref{leptones})  as~\footnote{Of course there is a similar Yukawa interaction with charged leptons obtained from Eq.~(\ref{yukawa}) by $Y_\n(y)\to Y_\e(y)$, $\n(x,y)\to \e(x,y)$ and $\widetilde H(x,y)\to H(x,y)$, by which zero modes of charged leptons get a mass after electroweak symmetry breaking.}
\be
\mathcal L_{\textrm{int}}=\int dy\, e^{-4ky}\ \bar\ell(x,y)\, Y_\n(y)\, \n(x,y)\cdot \widetilde H(x,y)+\textrm{h.c.}
\label{yukawa}
\ee
where $\widetilde H=i\sigma_2 H^*$ and $Y_\n$ is a matrix of 5D Yukawa couplings that can have both bulk  and brane contributions 
\be
Y_\n(y)=Y_\n^B+Y_\n^0\,\delta(y)+Y_\n^1\,\delta(y-y_1)\,.
\ee
The Higgs boson zero mode profile is given by
\be
h^{(0)}(y)=\frac{e^{a ky}}{N_h^{1/2}}\,,\qquad N_h=\frac{e^{2(a-1)ky_1}}{2(a-1)k}\,.
\label{higgs}
\ee
where $a>2$ in the RS metric to solve the hierarchy problem. 
After integration over $y$ one can write the neutrino mass matrix as
\be
m_\nu^{ij}=\frac{v}{\sqrt{N_h N_\ell^i N_\n^j}}\int_0^{y_1} (Y_\n)_{ij}\ e^{(a-c_{\ell}^i -c_\n^j) ky} 
\label{Dmass}
\ee

We will be using the fact that LH leptons should typically be leaning towards the UV brane in order to satisfy electroweak precision constraints, which implies $c^i_\ell>\frac{1}{2}$. We will also make the reasonable assumption that $$a>c_\ell^i+c_\n^j$$ as $a$ is constrained to be $a>2$ as we stated above. We will compute the neutrino mass matrix (\ref{Dmass}) separately for the cases of bulk and brane Yukawa couplings.

The neutrino mass matrix for bulk Yukawa coupling is given by
\be
m_\nu^{ij}=v\,Y^{ij}\left\{\begin{array}{ll}\epsilon^{c_\ell^i-1/2}\epsilon^{c_\n^j-1/2}& (c_\n^j>1/2)\\
\epsilon^{c_\ell^i-1/2}& (c_\n^j<1/2)
\end{array}
\right.
\label{bulkmass}
\ee
where
\be
\epsilon=e^{-k y_1}\simeq 10^{-15}
\label{epsilon}
\ee
in order to solve the hierarchy problem between $k\simeq M_P$ and the TeV scale,
and
\be
Y_{ij}=\frac{ \sqrt{2(a-1)|1-2 c_\ell^i|\,|1-2 c_\n^j|}}{a-c_\ell^i-c_\n^j}\ (Y_\n^B)_{ij}
\ee
To avoid suppressing the $\tau$ mass too much we need $\epsilon^{c_\ell-1/2}\gtrsim m_\tau/m_{t}\sim 10^{-2}$ (leading to $c_\ell\lesssim 0.63$) and therefore the case $c_\n<1/2$ in Eq.~(\ref{bulkmass}) is unrealistic. On the other hand the case $c_\n>1/2$ in Eq.~(\ref{bulkmass}) can easily describe the spectrum of neutrino masses provided that $c_\n\gtrsim 5/6$. However the neutrino spectrum predicted by Eq.~(\ref{bulkmass}) should be hierarchical unless some bulk symmetry (see section~\ref{flavor}) forces $c_\ell^i\equiv c_\ell$ and $c_\n^i\equiv c_\n$ for $\forall i$, which would render it anarchic as described by experimental data.

In the case of an IR brane Yukawa couplings
$
Y_\n=Y_\n^1\, \delta(y-y_1)
$
the result in Eq.~(\ref{bulkmass}) applies, except that
\be
Y_{ij}= \sqrt{2(a-1)|1-2 c_\ell^i|\,|1-2 c_\n^j|} \ (Y_\n^1)_{ij}\ ,
\ee
while for the case of a UV Yukawa coupling $Y_\n=Y_\n^0\,\delta(y)$ we get 
\be
m_\nu^{ij}=vY^{ij}\left\{\begin{array}{ll}\epsilon^{a-1}& (c_\n^j>1/2)\\
\epsilon^{a-1}\epsilon^{1/2-c^j_\n}& (c_\n^j<1/2)
\end{array}
\right.
\label{UVmass}
\ee
where
\be
Y_{ij}= \sqrt{2(a-1)|1-2 c_\ell^i|\,|1-2 c_\n^j|} \ (Y_\n^0)_{ij}\,,
\ee
At first this result seems promising as it is independent of the $c_\ell^i$ and hence allows for large mixing angles~\footnote{It was previously noted that for $c_\n+c_\ell>a$ the UV Yukawa couplings are naturally dominating over the IR ones, leading to flavor blindness of the 4D Yukawa couplings \cite{Agashe:2008fe}.}.
However, even for $c_\n>1/2$ this gives too small neutrino masses unless we consider $\epsilon\gtrsim 10^{-12}$, which implies  $k\lesssim 10^{15}$ GeV $\ll M_P$, in which case we do not solve the grand hierarchy problem (as in little RS models~\cite{Davoudiasl:2008hx}). As we only focus on theories solving the Planck/TeV hierarchy we will disregard this class of scenarios.

To summarize the case of Dirac neutrinos, its mass matrix and  mixing angles can be realistically described by different localization of leptonic fields in the bulk of AdS space provided there is a symmetry imposing $c_\ell^i\equiv c_\ell$ and $c_\n^i\equiv c_\n$ for $\forall i$ and both $c_\ell,c_\n>1/2$. We will come back to the issue of the bulk symmetry later on in this paper.

\section{Majorana neutrinos}
\label{Majorana}
The kinetic Lagrangian for one generation of RH neutrinos is given by Eq.~(\ref{A1}).
It is invariant under an $SU(2)_{\mathbb \n}$ global symmetry under which the neutrino transforms as a doublet with components $\n^{(1)}=(\n_L,\n_R)$, $\n^{(2)}=(\bar\n_R,-\bar\n_L)$~\footnote{This is sometimes referred to as a symplectic Majorana spinor.}. Only $\n\equiv\n^{(1)}$ will  couple to Higgs and leptons, and hence $SU(2)_{\mathbb \n}$ will be broken down to lepton number $\mathbb L\equiv U(1)_{\n}$ generated by $\sigma^3$.
The most general bulk mass term can be written in $SU(2)_{\mathbb \n}$ covariant form as $\bar \n^{(i)}\,\vec p\cdot \vec \sigma_{ij}\,\n^{(j)}$ with $\vec p$ a real three-vector \cite{Diego:2005mu} which also breaks $SU(2)_{\mathbb \n}$. Without loss of generality we will choose $\vec p=(c_M,0,c_\n)$, leading to 
\be
\mathcal L_{\rm mass}=\int
e^{-4ky}\,
M(y)\left( 
-c_\n\,\bar\n_R\n_L+{\rm h.c.}+\frac{1}{2}c_M [\n_L\n_L-\n_R\n_R+{\rm h.c.}]
\right)\,.
\label{A2}
\ee
Clearly $c_\n$ conserves $\mathbb L$, while $c_M$ breaks it. Further breaking of $\mathbb L$ can be introduced via the boundary conditions (BC), to be specified below. For the sake of generality, in Eq.~(\ref{A2}) we have allowed the bulk masses to depend on $y$. In the following we will only consider the constant case $M=k$, while the generic case for arbitrary $M(y)$ and $A(y)$ is worked out in App.~\ref{details}.

The expansion in modes reads for an arbitrary number of RH neutrinos
\be
\left(\begin{array}{c}
\n^a_L(x,y) 
\\
\n^a_R(x,y)
\end{array}
\right)
=\sum_n\left(\begin{array}{c}
\n^{(n)a}_L(x,y) 
\\
\n^{(n)a}_R(x,y)
\end{array}
\right)=\sum_n \left(\begin{array}{c}
\n^{(n)}(x)\ f^{(n)}_{\n^a_L}(y) 
\\
\bar\n^{(n)}(x)\ f^{(n)}_{\n^a_R}(y)
\end{array}
\right)
\ee
where the 4D Majorana spinor $(\n^{(n)}(x),\bar\n^{(n)}(x))$ has the 4D Majorana mass $m_n$. At this point we are only considering one generation of RH neutrinos and so the label $a$ will be removed.
Defining new wave functions  
\be
f_{\n_\chi}^{(n)}(y)=e^{2ky} \hat f_{\n_\chi}^{(n)}(y)\,,
\ee
we can rewrite the Dirac equation as 
\be
(m_n\, e^{2ky}\pm c_M\,k) {\hat f}^{(n)}_{\n_{L,R}}=(c_\n\,k\pm \partial_y)\hat f^n_{\n_{R,L}}\,.\label{Dirac2}
\ee
%
%
%

In order to obtain the BC, notice that variation of the action Eq.~(\ref{A1})  leads to 
$\n_L(0)
=\n_L(y_1)=0\,.
$
Including a 4D Majorana mass term for the non-vanishing field $\n_R$,
\be
\mathcal L_{\rm bd} = \left[\frac{n_0}{2}\,\bar\n_R\bar\n_R+{\rm h.c.}\right]_{y=0}
-\left[e^{-4ky_1}\,\frac{n_1}{2}\bar\n_R\bar\n_R+{\rm h.c.}\right]_{y=y_1}
\label{LB1}
\ee
(where we are considering real dimensionless numbers $n_i$) the BC's change to
\be
\n_L(0)+n_0{\bar\n}_R(0)=0\,,\qquad
\n_L(y_1)+n_1{\bar\n}_R(y_1)=0\,.
\label{BC1}
\ee
Clearly generic brane masses again violate $\mathbb L$. It is important to notice that besides $n_i=0$ also $n_i=\infty$  
conserve $\mathbb L$~\footnote{As pointed out in Ref.~\cite{Diego:2005mu} the most general BC can again be parametrized by two $SU(2)_{\mathbb L}$ unit vectors $\vec s_i$. The $\mathbb L$-conserving BC's correspond to the choices $\vec s_i=(0,0,\pm 1)$.}. 
%
%
%
%

%
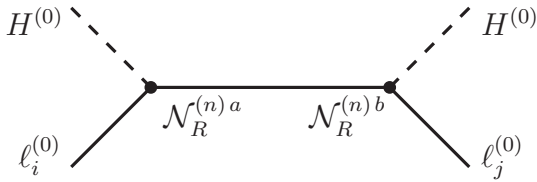
\begin{figure}[htb]
\begin{center}
\SetScale{1.2}
\begin{picture}(65,80)(35,-20)
\SetWidth{1.}
\Vertex(25,25){2}
\Vertex(100,25){2}
\DashLine(0,50)(25,25){4}
\Line(25,25)(0,0)
\Line(25,25)(100,25)
\DashLine(125,50)(100,25){4}
\Line(100,25)(125,0)
\put(35,15){$\n_R^{(n)\,a}$}
\put(90,15){$\n_R^{(n)\,b}$}
\put(-25,50){$H^{(0)}$}
\put(155,50){$H^{(0)}$}
\put(-20,0){$\ell^{(0)}_i$}
\put(155,0){$\ell^{(0)}_j$}
\end{picture}
\vspace{-1cm}
\end{center}
\caption{\it Diagram contributing to the Weinberg operator in the effective Lagrangian.}
\label{diagramas}
\end{figure}

In the general case of several generations of RH neutrinos we can always work in a basis where the bulk Dirac masses are diagonal. However the Majorana masses will, in general, be non-diagonal except if there is a bulk symmetry implying diagonal masses. 
Moreover in this work we will always impose reality of both bulk and brane mass (as well as Yukawa) matrices~\footnote{Complex entries would introduce additional CP violating phases which we are not considering in this work.}. We will consider the 5D Yukawa interaction between a bulk Higgs field $H(x,y)$, the leptons $\ell^{\,i}(x,y)$ and $N_\n$ RH neutrinos~\footnote{Since for the time being we are keeping the number of RH neutrinos $N_\n$ arbitrary we will label them with indices $a,b$ as opposed to $i,j,\dots$ which is used for the three generations of charged leptons.} $\n^a(x,y)$ as in Eq.~(\ref{yukawa}) and will integrate out the whole tower of RH neutrinos 
as in Fig.~\ref{diagramas}, giving rise to the 4D effective operator
\be
\mathcal L_W=c_W^{ij}  \left[\bar\ell^{(0)}_i(x)\cdot\tilde H^{(0)}(x)\right]\, \left[\tilde H^{T(0)}(x)\cdot \ell^{\,c(0)}_j(x)\right]+\textrm{h.c.}
\label{WO}
\ee
where 
we denote by $H^{(n)}(x)$ and $\ell^{(n)}_i(x)$ the normalized 4D modes of the Higgs and leptons doublets.
To proceed we calculate the 5D propagator at zero momentum,
\be
G_{RR}^{ab}(y,y')=\sum_n \frac{\hat f_{\n_R^a}^{(n)}(y)\hat f^{(n)}_{\n^b_R}(y')}{m_n}
\label{def}
\ee
a quantity which is expected to vanish whenever $\mathbb L$ is a good symmetry of the theory. Then the coefficient $c_W^{ij}$, with mass dimension $-1$, is given by
\be
c_W^{ij}=\int dy\, dy'\ Y^{ia}_{\n}(y)\, Y^{jb}_{\n}(y')\ h^{(0)}(y)\,h^{(0)}(y')\
 \hat f^{(0)}_{\ell^i_L}(y)\,\hat f^{(0)}_{\ell_L^j}(y')\ G_{RR}^{ab}(y,y')
\label{cW}
\ee
where $h^{(0)}(y)$ and $\ell^{(0)}_i(y)$ are given in Eqs.~(\ref{higgs}) and (\ref{fermion}), respectively,
and the neutrino mass matrix by
\be
m_\nu^{ij}=c_W^{ij}v^2 .
\label{mass}
\ee
 %
In the following we will investigate the implications for lepton physics in two particularly interesting cases. In Sec.~\ref{cMnonzero} the general case of $c_M\neq 0$ will be studied, while in Sec.~\ref{cMzero} we will consider $c_M=0$.


\subsection{The general case $c_M\neq 0$.}
\label{cMnonzero}

Let us first study the case where $\mathbb L$ is violated in the bulk (and possibly on the branes). As it turns out one only needs to introduce one generation of 5D RH neutrinos to give Majorana masses to the three left-handed neutrinos. The propagator (\ref{def}) is then given by
\be
G_{RR}(y,y')= \\\frac{c_M}{\ceff} \frac{
 [c_0 \sinh(\ceff Q_m)-\ceff\cosh(\ceff Q_m)  ][c_1 \sinh(\ceff Q_M)+ \ceff\cosh(\ceff Q_M)]} 
 {  (c_0c_1-\ceff^2)\sinh(\ceff Q_1) + \ceff(c_0-c_1)\cosh(\ceff Q_1)}\label{equ_GRR}
\ee
where we have defined $\ceff=\sqrt{c_\n^2+c_M^2}$, $c_i=c_Mn_i+c_\n$ as well as 
$Q_m=k\min(y,y')$, $Q_M=k\,y_1-k\max(y,y')$ and $Q_1=ky_1$.
Details of the calculation for $G_{RR}$ can be found in App.~\ref{details} (see also Ref.~\cite{Watanabe:2010cy}).

For a bulk Yukawa coupling $Y_\n(y)=Y_\n^B$ the coefficient of the Weinberg operator (\ref{WO}) can then be written as
\bea
c_W ^{ij}&=& K_{ij}\epsilon^{-1}\left[
f_2^{ij}(\ceff) + 
f_1^{ij}(\ceff) \epsilon^{a_j+c_\nu} + 
f_0^{ij}(\ceff) \epsilon^{a_i+a_j} \right.\nn\\
&+&\left.
f_2^{ij}(-\ceff) \epsilon^{2\ceff} + 
f_1^{ji}(\ceff) \epsilon^{a_i+\ceff} + f_0^{ij}(-\ceff) \epsilon^{a_i+a_j+2\ceff } \right]
\eea
where 
%
%
%
we have defined
\bea
f_0^{ij}(\ceff )&=&-2\ceff (c_1+\ceff )(a_i+\ceff  )(a_j+\ceff )(a_i+a_j+c_0-\ceff )\\
f_1^{ij}(\ceff )&=&4\ceff ^2(a_i+a_j)(a_i+c_1)(a_j+c_0)\\
f_2^{ij}(\ceff )&=&2\ceff (c_0-\ceff )(a_i-\ceff )(a_j-\ceff )(a_i+a_j+c_1+\ceff )
\eea
with $a_i\equiv a-c_\ell^i$, 
\be
K_{ij}=\frac{c_M(a-1)}{\ceff }
 \frac{\left[
 (c_0-c_\nu)(c_1+c_\nu)-(c_0+c_\nu)(c_1-c_\nu)\epsilon^{2c_\nu}
 %
 %
 \right]^{-1} } {(a_i+a_j)(a_i-\ceff) (a_j-\ceff ) }Y_iY_j
\ee
and
\be
Y_{i}=
\sqrt{\frac{(2c^i_\ell-1)\epsilon^{2c^i_\ell-1}}{(1-\epsilon^{2c^i_\ell-1})}}\frac{(Y_\n^{B})_i}{(a_i+\ceff)}
\label{Yukawabulk}
\ee
The leading term will be provided by the term proportional to $f_2(\ceff)$, resulting in
\be
c_W^{ij}=\frac{2(a-1)c_M
\epsilon^{-1}}{
(c_1+c_\nu)
}
\frac{(a_i+a_j+c_1+\ceff)}{(a_i+a_j)}
Y_iY_j
\label{cWbulk}
\ee
Notice that generically (for arbitrary values of the constants $a_i$) the rank of the matrix $c_W$ is equal to three [$r(c_W)=3$] and, unlike in the 4D seesaw mechanism, one does not need to introduce several RH neutrinos to ensure that no more than one LH neutrino is massless.

Using the fact that LH leptons should typically be leaning towards the UV brane ($c^i_\ell>\frac{1}{2}$) and assuming $Y_\n^{B}=\mathcal O(1)/\sqrt{k}$, one can estimate from (\ref{Yukawabulk}) and (\ref{cWbulk}) the order of magnitude of the neutrino mass matrix as
\be
m_\nu^{ij}\simeq \mathcal O(1)^{ij}\ \frac{v^2}{\epsilon k}\,\epsilon^{c_\ell^i+c_\ell^j-1}\simeq \, \epsilon^{c_\ell^i+c_\ell^j-1}\,10^{11}\ \textrm{eV}
\label{masa}
\ee
Assuming now that $c_\ell^i\equiv c_\ell$ ($\forall i$) to describe the large neutrino mixing and using the bound $\epsilon^{2 c_\ell-1}\gtrsim 10^{-4}$ imposed from the $\tau$-mass we are led to the condition $m_\nu\gtrsim 10$ MeV, in flagrant conflict with experimental data on neutrino masses~\footnote{\label{comment}Of course this case has another flaw. If $c_\ell^i\equiv c_\ell$ ($\forall i$) it turns out that the rank of the matrix $c_W$ is $r(c_W)=1$ [as it can easily be checked from Eq.~(\ref{cWbulk})] and the two light neutrinos are massless. This problem could be fixed if the symmetry enforcing equality of all $c_\ell^i$ is approximate and $c_\ell^i\simeq c_\ell$. However this case is unrealistic anyway as we have noticed because the third generation neutrino is too heavy and we will disregard it.}. In the next section we will work out the particular case $c_M=0$ where the behaviour (\ref{masa}) appears modified by an extra suppression factor and one can then overcome the above problem and decouple the behavior of the neutrino mass matrix from that of the charged leptons.

For an IR localized Yukawa coupling $Y_\n(y)=Y^1_\n\,\delta(y-y_1)$ and using $G_{RR}(y_1,y_1)=c_M/(c_1+c_\nu)$ the coefficient of the Weinberg operator (\ref{WO}) can then be written as
\be
c_W^{ij}=\frac{2(a-1)c_M
\epsilon^{-1}}{
(c_1+c_\nu)
}
Y_iY_j
\label{cWIR}
\ee
where
\be
Y_{i}=
\sqrt{\frac{(2c^i_\ell-1)\epsilon^{2c^i_\ell-1}}{(1-\epsilon^{2c^i_\ell-1})}}\,(Y_\n^{1})_i
\label{YukawaIR}
\ee
so that we obtain similar conclusions to the case of a bulk Yukawa coupling. Similarly, for a UV localized Yukawa coupling  $Y_\n(y)=Y_\n^0\,\delta(y)$ and using $G_{RR}(0,0)=c_M/(c_\nu-c_0)$ one can write
\be
c_W^{ij}=\frac{2(a-1)c_M
\epsilon^{-1}}{
(c_\nu-c_0)
}\, \epsilon^{2a-1}
Y_iY_j
\label{cWUV}
\ee
where
\be
Y_{i}=
\sqrt{\frac{(2c^i_\ell-1)}{(1-\epsilon^{2c^i_\ell-1})}}\,(Y_\n^{0})_i
\label{YukawaUV}
\ee
which gives too small values for the neutrino mass $m_\nu^{ij}$. Also notice that in the case where Yukawa couplings are exclusively localized towards one of the branes, $r(c_W)=1$ and the comments in footnote~\ref{comment} do apply.

To summarize the case where $\mathbb L$ is broken in the bulk, i.e.~$c_M\neq 0$: \textbf{i)} For the case of Yukawas in the bulk and/or localized on the IR boundary, the predicted neutrino masses are in the MeV range and therefore excluded by experimental data. \textbf{ii)} On the contrary for Yukawas localized on the UV boundary the predicted neutrino masses are smaller than experimental data by many orders of magnitude. We will see that these problems can be solved if the theory conserves lepton number in the bulk but breaks it on the branes.


\subsection{The case $c_M=0$}
\label{cMzero}

If the 5D theory conserves lepton number $\mathbb L$, $c_M=0$ and the bulk Majorana mass term vanishes. In this case considering an arbitrary number $N_\n$ of RH Majorana neutrinos one obtains the RR-propagator matrix as
\be
G_{RR}(y,y')=e^{-c_\n ky}\left(e^{-c_\n ky_1}n_1e^{-c_\n ky_1}-n_0\right)^{-1}e^{-c_\n ky'} \label{equ_GRR_cMzero}
\ee
where $c_\n^{ab}=\delta^{ab}c_\n^b$ and $n_i^{ab}$ are arbitrary (symmetric, dimensionless) brane Majorana mass matrices appearing in Eq.~(\ref{LB1}). 

For a bulk Yukawa coupling $Y_\n(y)=Y_\n^B$ we write the coefficient $c_W^{ij}$ of the Weinberg operator, Eq.~(\ref{WO}), as
\be
c_W=Y\left(\epsilon^{c_\n}\,n_1\,\epsilon^{c_\n}-n_0\right)^{-1}Y^T
\ee
where %
\be
Y_{ia}= Y^{B}_{ia}\ 
\frac{\epsilon^{c^a_\n+c^i_{\ell}-1}-\epsilon^{a-1}}{a-c^a_\n-c^i_{\ell}}\
\sqrt{\frac{{2(a-1)(2c_\ell^i-1)}}{{1-\epsilon^{2c_\ell^i-1}}}}
\ee

There are different regimes, depending on the values of the numbers $c_\n$, $c_\ell$ and $a$. 
For $2<a<c_\n+c_\ell$ we obtain a neutrino mass scale that is suppressed at least as 
$
m_\nu\sim v^2 \epsilon^{2(a-1)}/k\,<10^{-34}\ {\rm eV}
$ and hence~\footnote{In fact it has been pointed out in Ref.~\cite{Agashe:2008fe} that this case could be potentially interesting to explain the anarchic structure in the neutrino sector. However, as already noted there in warped space the Majorana case is unrealistic due to too large suppression of the neutrino mass scale.}
we are lead to consider the case $c_\n+c_\ell<a$. In this case assuming that the left-handed leptons lean towards the UV boundary and $c_\n>0$ one can simplify the expression for $c_W$ as
\begin{equation}
c_W=-Y\,n_0^{-1}\,Y^T\label{cWbueno}
\end{equation}
and the neutrino Majorana mass matrix
\be
m_{\nu}^{ij}\sim \mathcal O(1)^{ij}_{ab}\ \frac{v^2}{\epsilon\, k}\,\epsilon^{c_\ell^i+c_\ell^j-1}\ \epsilon^{c_\n^a}(n_0)^{-1}_{ab}\epsilon^{c_\n^b}
\label{masa0}
\ee
contains an extra suppression factor $\epsilon^{c_\n}(n_0)^{-1}\epsilon^{c_\n}
$ with respect to the $c_M\neq 0$ case, Eq.~(\ref{masa}), which makes (\ref{masa0}) consistent with the charged lepton spectrum \cite{Huber:2003sf}.

For Yukawa couplings localized on the IR, $Y_\n(y)=Y_\n^1\,\delta(y-y_1)$, or UV, $Y_\n(y)=Y_\n^0\,\delta(y)$, boundaries the result (\ref{cWbueno}) for the coefficients $c_W^{ij}$ hold with the corresponding respective definitions
\begin{eqnarray}
Y_{ia}&=& Y^{1}_{ia}\ 
\epsilon^{c^a_\n+c^i_{\ell}-1}\
\sqrt{\frac{{2(a-1)(2c_\ell^i-1)}}{{1-\epsilon^{2c_\ell^i-1}}}}
\\
Y_{ia}&=& Y^{0}_{ia}\ 
\epsilon^{a-1}\
\sqrt{\frac{{2(a-1)(2c_\ell^i-1)}}{{1-\epsilon^{2c_\ell^i-1}}}}
\end{eqnarray}
 We see that the case of an IR localized Yukawa coupling gives a similar expression for the neutrino mass as the bulk Yukawas coupling, as in the previous cases, while a UV localized Yukawa coupling provides unrealistic results which translate into too small values of the neutrino mass.

To summarize this section we can see that the cases of a bulk or IR localized Yukawa coupling matrix ($Y_{ia}$) can provide a convenient description of the Majorana neutrino mass matrix. However the largeness of the neutrino mixing angles requires that $c_\ell^i=c_\ell$ and $c_\n^a=c_\n$ for $\forall i,a$, which requires a (gauge) bulk symmetry as we will discuss in the next section. In that case the general theorem on rank of matrix product implies that $r(c_W)\leq \min \{r(n_0^{-1}),r(Y)\}$ which in turn implies that to give masses to the three LH neutrinos we need $N_\n\geq 3$ and moreover $r(n_0^{-1})\geq 3$ and $r(Y)=3$ (i.e.~full rank), as in the cases we will consider in Sec.~\ref{flavor} where we will fix $N_\n=3$.

\section{Flavor from Wilson Lines}
\label{flavor}
As we have seen in the previous sections the anarchic neutrino spectrum requires (unlike the hierarchical quark structure) a symmetry in the bulk.  
One possibility is that this symmetry is gauged in 5D. According to the AdS/CFT correspondence a 5D local  symmetry implies the existence of an exact (or spontaneously broken) global symmetry of the 4D dual theory. 
Moreover, this symmetry is expected to be rather large in order to ensure degeneracy amongst the various Dirac bulk masses,  but needs to be broken to a sufficiently small subgroup at the boundaries in order to allow for nontrivial Yukawa couplings. There will thus in general be a nontrivial coset in which some of the fifth components of the gauge fields, $A_5$, can acquire VEVs.
Hence, it is natural to ask
 if the mixing in the lepton sector could come from nontrivial Wilson lines or, equivalently, nonzero VEVs for $A_5$. 

If we start with a group $G$ in the bulk and break it to subgroups $H_0$ and $H_1$ at the boundaries, the theory has zero modes for $A_\mu\in H_0\cap H_1\equiv H$ and $A_5\in K_0\cap K_1\equiv K$ with $K_i\equiv G/H_i$. The profiles for the latter are given by 
\be
A_5(y)=\theta e^{2ky}
\ee
where $\theta$ is a matrix in the above coset. Each coset is spanned by the generators
\be
K_i=\left\{T\in \mathcal G|\tr T\mathcal H_i=0\right\}
\ee
where we denote by $\mathcal G$ ($\mathcal H_i$) the Lie algebra of $G$ ($H_i$). This is a system of $\dim H_i$ linear equations defining a $\dim G-\dim H_i$ linear subspace. $K=K_0\cap K_1$ is the space of possible zero modes for $A_5$. 

The zero mode for $A_5$ has no potential at tree level and all configurations are degenerate, but can get a VEV through radiative corrections although here we will not specify the possible dynamics leading to different configurations.  One can transform away $A_5$ by a gauge transformation.
\be
\Lambda(y)=i \int_y^{y_1} A_5(y)
\ee
The bulk action is left unchanged because of gauge invariance, provided we also transform all fields charged under $G$. 
This changes the UV BC of all fields transforming under the gauge group. In particular 
one has to make the replacement 
\be
\psi(0)\to e^{i\Lambda_0}\psi(0)\,,\qquad \Lambda_0\equiv\Lambda(0)=\theta\int_0^{y_1} e^{2ky}
\label{replacement}
\ee 
taken in the appropriate representation. The UV boundary condition for the RH neutrinos in Eq.~(\ref{BC1}) correspondingly changes as
\be
n_0\to e^{-i\Lambda_0}n_0e^{-i\Lambda_0^T}\,.
\label{n0transformed}
\ee

\subsection{Choice of gauge group}

We will consider the case of three generations of right handed neutrinos. The free 5D action, including bulk kinetic terms,
is invariant under $U(3)_\e\otimes U(3)_\ell\otimes U(3)_\n$. The bulk gauge group $G$ should be a subgroup of the latter.
In choosing  $G$ one should take into account the following requirements:
\begin{itemize}
\item
$G$ needs to be large enough such that the breaking $G\to H$ allows for nontrivial Wilson lines.
\item
In general, the larger $G$, the more the theory will be protected from flavor changing neutral currents (FCNC).
\item
$G$ should ensure degeneracy of the $c_\ell$, but allow for non-degenerate $c_\e$.
\item
$G$ should not be so large such that the breaking $G\to H$ leaves over unwanted zero modes for $A_\mu$ (4D gauge symmetries). 
\end{itemize} 
Since we would like to avoid hierarchical mixing, a natural choice is to take $G$ to include $U(3)_\ell$ and to write a Wilson line to rotate RH neutrinos, we should also have $G\supset U(3)_\n$~\footnote{An even more minimal choice would be $U(3)_{\ell+\n}$.}. As for the charged leptons we would like to be able to write different $c_{\e^i}$, so there should not be any nonabelian transformation on the charged leptons. 
A common charged lepton mass term will leave $U(3)_\e$ unbroken while different masses will break it to $\otimes_iU(1)_{\e^i}\subset U(3)_\e$. The maximal group that allows to generate different charged lepton masses is thus $\otimes_i U(1)_{\e^i}$.

To summarize the simplest bulk gauge group is
\be
G=U(3)_\ell\otimes U(3)_\n\otimes_{i}U(1)_{\e^i}
 \label{simplest}
 \ee
where $i=1,2,3$ runs over the number of generations. In the following we are denoting the $U(3)_\psi$ generators by $\lambda^\alpha_{\psi}$ ($\alpha=0\dots 8$) where $\lambda^\alpha$ are Gell-Mann matrices (normalized to $\frac{1}{2}$) for $\alpha=1\dots 8$ and $\lambda^0=\diag (1,1,1)/\sqrt{6}$. The last factor $\otimes_iU(1)_{\e^i}$ can be normalized such that it is spanned by $\{\lambda_\e^3,\lambda_\e^8,\lambda_\e^0\}$.
The first and second factors are consistent with constant (in generation space) $c_\ell$ and $c_\n$ while the last one introduces differents $c_{\e^i}$ as it is required to describe the charged lepton spectrum. On the other hand bulk Yukawas and the bulk Majorana mass vanish
 \be
Y^B_\n= Y^B_\e=c_M=0 \ .
 \ee

On the IR boundary the group $G$ can be broken to
\be
H_1=\otimes_iU(1)_{(\ell+ \e+ \n)^i}=\{\lambda_\e^3+ \lambda_\ell^3+ \lambda_\n^3,\ \lambda_\e^8+ \lambda_\ell^8+ \lambda_\n^8,\ \lambda_\e^0+ \lambda_\ell^0+ \lambda_\n^0\}
\label{IRH}
\ee
i.e.``$\mathbb L$epton family number" which is consistent with diagonal Yukawas 
\be
Y_{\n ij}^1=Y_{\n i}^1\;\delta_{ij},\quad Y_{\e ij}^1=Y_{\e i}^1\;\delta_{ij}
\ee
and forbids a Majorana mass $(n_1=0)$. Finally on the UV boundary the gauge group $H_0$ is largely arbitrary. To allow for a Majorana mass $n_0$ it should not contain $U(1)_\n$, but it can in general contain a subgroup of $SU(3)_\n$. Possible choices are $SO(3)_\n$, $U(1)_\n\subset SU(3)_\n$ or even completely broken $U(3)_\n$.   For definiteness we will choose here the second possibility with the generator being $\lambda_\n^1$.
The part of $G$ acting on the doublets and charged leptons may be left unbroken to avoid additional Wilson line moduli in those sectors.
We thus take the surviving group to be
\be
H_0=U(3)_\ell\otimes U(1)_{\lambda^1_\n}\otimes_i U(1)_{\e^i}
\label{UVH}
\ee
%
This leads to zero UV brane Yukawa couplings
\be
Y^0_\n=Y^0_\e=0
\ee
while the Majorana mass matrix $n_0$ has to fulfill
\be
n_0\lambda_\n^1+\left(\lambda^1_\n\right)^T n_0=0\ .
\label{n0anti}
\ee
in order to be $H_0$ invariant.
One can easily check that $H=H_0\cap H_1=\varnothing$ while 
\be
K=\left\{ \lambda_\n^{2,4,5,6,7}\right\}
\label{coset}
\ee

 In this case and using the previously obtained structure of Yukawas and Majorana masses one can write
\be
c_W=-Y^Te^{i\Lambda_0^T}n_0^{-1}e^{i\Lambda_0}Y
\label{nuestro} 
\ee
where the Yukawa coupling is given by
\be
Y_{ij}= {Y^{1}_\n}_{ij}\ 
\epsilon^{c_\n+c_{\ell}-1}\
\sqrt{\frac{{2(a-1)(2c_\ell-1)}}{{1-\epsilon^{2c_\ell-1}}}}
\ee

A more minimal option is $G= U(3)_{\ell+\n}\otimes_iU(1)_{\e^i}$ with the boundary groups being $H_0=U(1)_{\ell+ \n}\otimes_i U(1)_{\e}^i$ and $H_1$ given by (\ref{IRH}). Still $H=\varnothing$ and $K$ is spanned by $\left\{ \lambda_{\ell+\n}^{2,4,5,6,7}\right\}$. In this case $Y^B_\n$ is nonzero and proportional to the unit matrix while Eq.~(\ref{n0anti}) holds for $\lambda^1_{\ell+\n}$ being a generator of $SU(3)_{\ell+\n}$. As we have seen in the previous sections that bulk and IR Yukawas provide similar contributions to the neutrino mass matrix, while UV Yukawas lead to subleading contributions, the phenomenology of both models should be very similar.

Let us finally comment that other choices of the subgroup of $U(3)_\n$ in $H_0$ would lead to different class of models, with different coset spaces $\mathcal K$ and different Majorana mass matrices $n_0$. In particular the choice $SO(3)_\n\subset H_0$ leads to $n_0=n_0^1 \diag(1,1,1)$ and $\mathcal K=\{\lambda^{1,4,6} \}$ while the choice of $U(3)_\n$ completely broken $\varnothing_\n \subset H_0$ leads to $n_0= \diag(n_0^1,n_0^2,n_0^3)$ and $\mathcal K=\{\lambda^{1,2,4,5,6,7} \}$. As a working example we will analyze in the next section the case where $U(1)_{\lambda_\n^1}\subset H_0$ but keeping in mind that other cases are possible and could give rise to a different phenomenology.

\subsection{A Wilson line model for $U_{PMNS}$}

The most general solution to Eq.~(\ref{n0anti}) is given by
\be
n_0=\diag\left(\, n_0^1,\, -n_0^1,\,n_0^3\, \right)
\ee
with $n_0^1$ and $n_0^3$ arbitrary numbers. Complex entries will result in additional phases in the PMNS matrix and for simplicity we will take $n_0^{1,3}$ to be real. For the same reason we will only consider the subspace
\be
 \left\{ \lambda_\n^{2,5,7}\right\}\subset K
\ee
which coincides with the generators of the group $SO(3)_\n\subset SU(3)_\n$ and leads to a real WL~\footnote{Of course the question which of the $A_5$ field directions will acquire a VEV is a dynamical problem which should be attacked by considering the one-loop Coleman-Weinberg potential, by applying for instance the general methods of Ref.~\cite{vonGersdorff:2002as} to warped space. As field directions along $\mathcal K$ are flat at the tree-level (as we are considering here) and a full one-loop analysis is outside the scope of the present paper we will just assume in the rest of this section that only $\langle A_5^{2,5,7}\rangle\neq 0$ and evaluate the region in the parameter space allowed by experimental data. If the field directions  $\langle A_5^{4,6}\rangle$ turn out to acquire a VEV then the rest of this section could be easily modified to cope with it. 
}.
For convenience we define the ratio
\be
y_3=2\,\frac{n_0^3}{n_0^1}
\ee
such that the inverse matrix appearing in Eq.~(\ref{nuestro}) is proportional to 
\be
n_0^{-1}\propto\diag\left(\frac{y_3}{2},-\frac{y_3}{2},1\right)
\label{n0}
\ee
The Yukawa matrix is diagonal because of the symmetry $H_1$ in Eq.~(\ref{IRH}),  and can be paremetrized as
\be
Y\propto\diag(y_1,y_2,1),
\label{diag-yukawa}
\ee 
where we have factored out a global constant and have then normalized its last entry to $1$. On the other hand using the fact that the coset $K\supset SO(3)_\n$ 
we can choose the Wilson Line  along $SO(3)$ as:
\be
\Lambda_0(b_k)=\left(\begin{array}{ccc}
0&-b_3i&b_2i\\b_3i & 0& -b_1i\\-b_2i&b_1i&0\\
\end{array}\right)
\ee
which is a linear combination of the generators $\lambda_{2,5,7}$ of $SU(3)$ which span $SO(3)\subset SU(3)$, where $b_k$ are real parameters. 
The parameters $b_i$ are thus periodic variables, and without loss of generality we can take them to satisfy $\sum_i b_i^2\leq \pi^2$.
We can now compute the WL $e^{i\Lambda_0}$ as a matrix with entries depending on the parameters $b_k$ which 
in the following we will denote as 
\be
\widehat U(b_k)\equiv e^{i\Lambda_0(b_k)}.
\ee
%
%
%
%

We can now use Eq.~(\ref{nuestro}) to write the neutrino mass matrix as
\be
m_\nu(b_k,y_k)\propto \diag(y_1,y_2,1)\cdot \widehat U^T(b_k)\cdot \diag\left(\frac{y_3}{2},-\frac{y_3}{2},1\right)\cdot \widehat U(b_k)\cdot \diag(y_1.y_2,1)
\label{neutrinomass}
\ee
where the dot indicates matrix product. 
The proportionality constant is given (up to $\mathcal O(1)$ numbers) by 
$\epsilon^{2c_\ell-1+2c_\n} v^2/\epsilon\, k$ which, using the value of $c_\ell$ required to fix the $\tau$ mass, becomes
$\sim  \epsilon^{2 c_\n} 10^6$ eV. Thus the proportionality constant is entirely controlled by $c_\n$ and will be consider  as a free parameter of the theory. 

Once we have determined the neutrino mass matrix (\ref{neutrinomass}) we want to fit the experimental data  consisting of the mass engenvalues $m_i(b_k,y_k)$ and the mixing angles of the matrix which diagonalizes the mass matrix. We parametrize the mixing angles as
\be
U(b_k,y_k)=\left(\begin{array}{ccc}
c_{12} c_{13}& s_{12} c_{13} & s_{13}\\
-s_{12} c_{23} - c_{12} s_{13} s_{23}& 
   c_{12} c_{23} - s_{12} s_{13} s_{23}&
   c_{13} s_{23}\\
   s_{12} s_{23} - c_{12} s_{13} c_{23} & -s_{12} s_{13} c_{23} - c_{12} s_{23} & c_{13} c_{23}
\end{array}
\right)
\ee
where we have neglected $CP$ violation in the leptonic sector (as there are no experimental data on it) and we are using the notation $s_{ij}(b_k,y_k)=\sin\theta_{ij}(b_k,y_k)$ for $ij=13$, 23 and  12. For the mixing angles we will use the following experimental values~\cite{Nakamura:2010zzi}
\begin{eqnarray}
(s_{13}^2)_{exp}&=&  0.023\pm 0.004\nonumber\\
(s_{12}^2)_{exp}&=&0.312\pm 0.016  \nonumber\\
(s_{23}^2)_{exp}&=&  0.52\pm 0.06
\label{mixing}
\end{eqnarray}
For the neutrino mass eigenvalues, for both normal ($|m_1|<|m_2|<|m_3|$) or inverted ($|m_3|<|m_1|< |m_2|$) ordering, the experimental data require
\be
r=\frac{m_2^2-m_1^2}{|m_3^2-m_2^2|},\quad r_{exp}=\frac{\Delta m^2_{\odot}}{\Delta m^2_{A}}=0.0312\pm 0.0018
\label{r}
\ee
where $\Delta m^2_{A}=(2.32^{+0.12}_{-0.08})\times 10^{-3} \textrm{ eV}^2$ and $\Delta m^2_{\odot}=(7.50\pm 0.20)\times 10^{-5}\textrm{ eV}^2$ are the atmospheric and solar neutrino squared mass differences respectively.
The reason we only consider here the ratio $r$ is that the overall normalization of the mass matrix can easily be adjusted to account for the correct absolute neutrino mass scale.
We now want to diagonalize the neutrino mass (\ref{neutrinomass}) and make a fit to the mass eigenvalues and mixing angles by means of the $\chi^2$ function
\be
\chi^2(b_k,y_k)=\sum_{ij}\left(\frac{s_{ij}^2(b_k,y_k)-(s_{ij}^2)_{exp}}{\Delta s_{ij}^2}\right)^2+\left(\frac{r(b_k,y_k)-r_{exp}}{\Delta r}\right)^2
\label{chisquare}
\ee

Notice that there is a qualitative difference between the parameters $b_k$ and $y_k$ as the former define the Wilson Line and should arise from some bulk dynamics while the latter are external parameters which should be $\mathcal O(1)$ as dictated by the anarchy assumption. Notice also that in the case where the Yukawa matrix is proportional to the identity (i.e. $y_1=y_2=1$) then $r=0$. In that case the best (very bad) fit corresponds to $U(b_k,y_k)=U_{PMNS}$ which happens for $(b_1^0,b_2^0,b_3^0)=(0.83,0.10,0.62)$ and $\chi^2_{min}=276$. A better fit requires departure from 1 of $y_1$ and/or $y_2$ to cope with the experimental value of the $r$-parameter. 

\begin{figure}[htb]
\begin{center}
\includegraphics[width=0.32\textwidth]{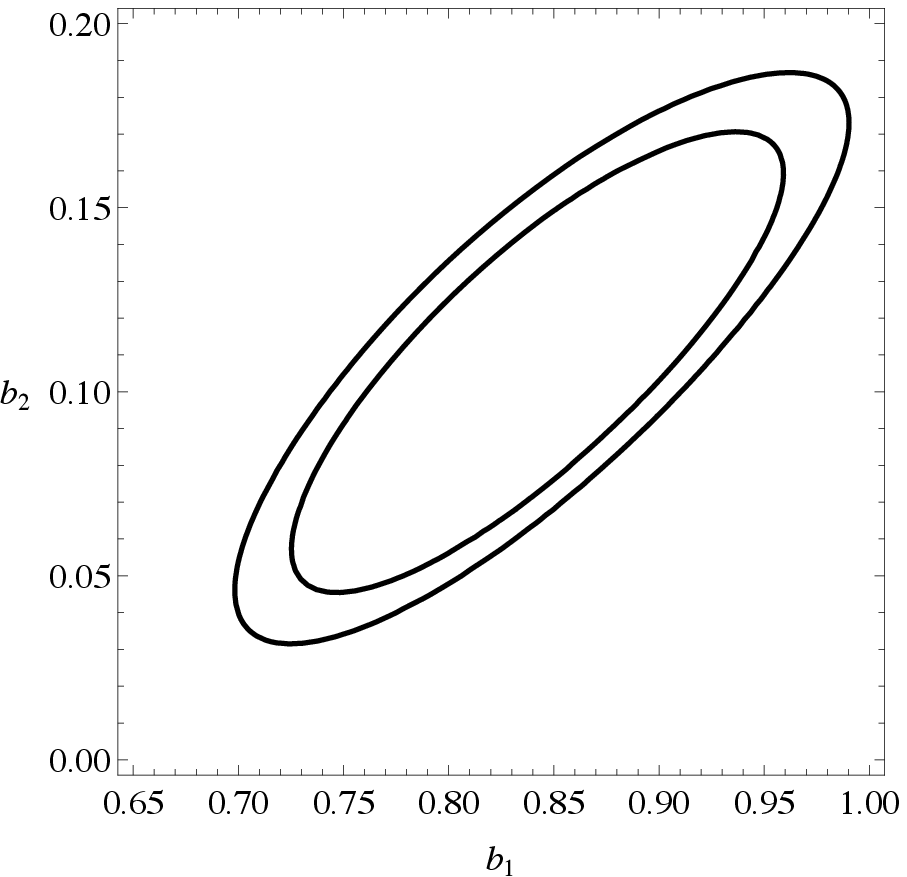}
\includegraphics[width=0.32\textwidth]{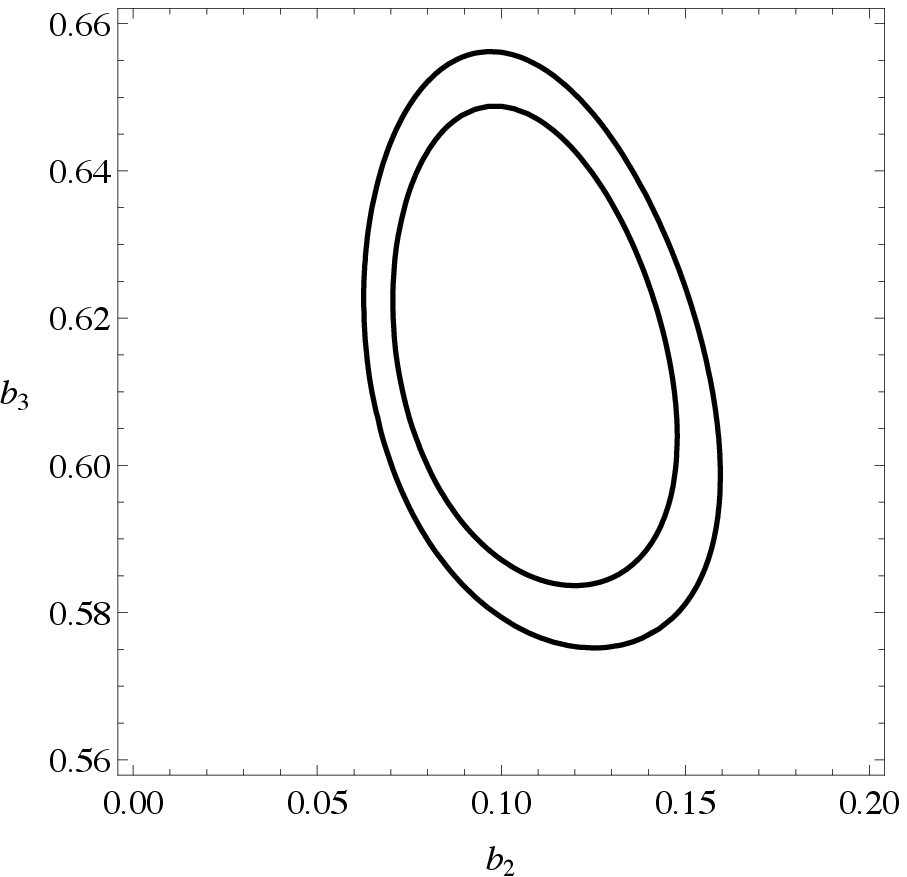}
\includegraphics[width=0.32\textwidth]{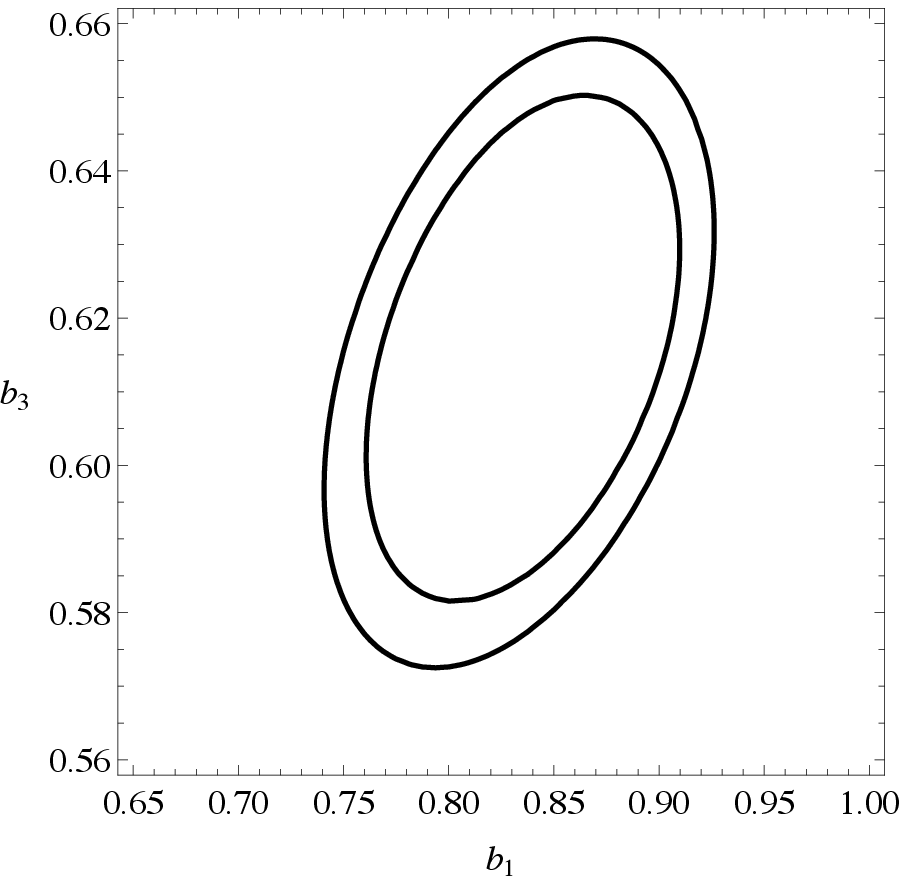}
\end{center}
\caption{\it 95\% CL (inside the inner ellipse) and 99\% CL (inside the outer ellipse) for $y_k=(0.90,0.95,0.90)$ in the plane $(b_1,b_2)$ with $b_3=b_3^0$ [left panel],  $(b_2,b_3)$ with $b_1=b_1^0$ [middle panel] and $(b_1,b_3)$ with $b_2=b_2^0$ [right panel].}
\label{b}
\end{figure}

A very simple example is the case where the Yukawa and Majorana mass matrices are such that $y_k\simeq 1$. Consider for instance the case where $y_k=(0.90,0.95,0.90)$ such that $\chi^2_{min}\simeq 0$ for $b_k^0=(0.84,0.11,0.62)$. 
In this case the neutrino mass spectrum at the best fit value is given by $m_i\simeq (0.022,0.024,0.055)$ eV which has a normal hierarchy. In Fig.~\ref{b} the 95\% CL and 99\% CL regions are shown in the $(b_1,b_2)$ plane for $b_3=b_3^0$ (left panel), in the $(b_2,b_3)$ plane for $b_1=b_1^0$ (middle panel) and in the $(b_1,b_3)$ plane for $b_2=b_2^0$ (right panel). 
Different plots in Fig.~\ref{b} are useful to measure the available region in the space $(b_1,b_2,b_3)$ which is consistent with experimental data and with fixed values of Yukawas and Majorana mass matrix entries.  

\begin{figure}[htb]
\begin{center}
\includegraphics[width=0.32\textwidth]{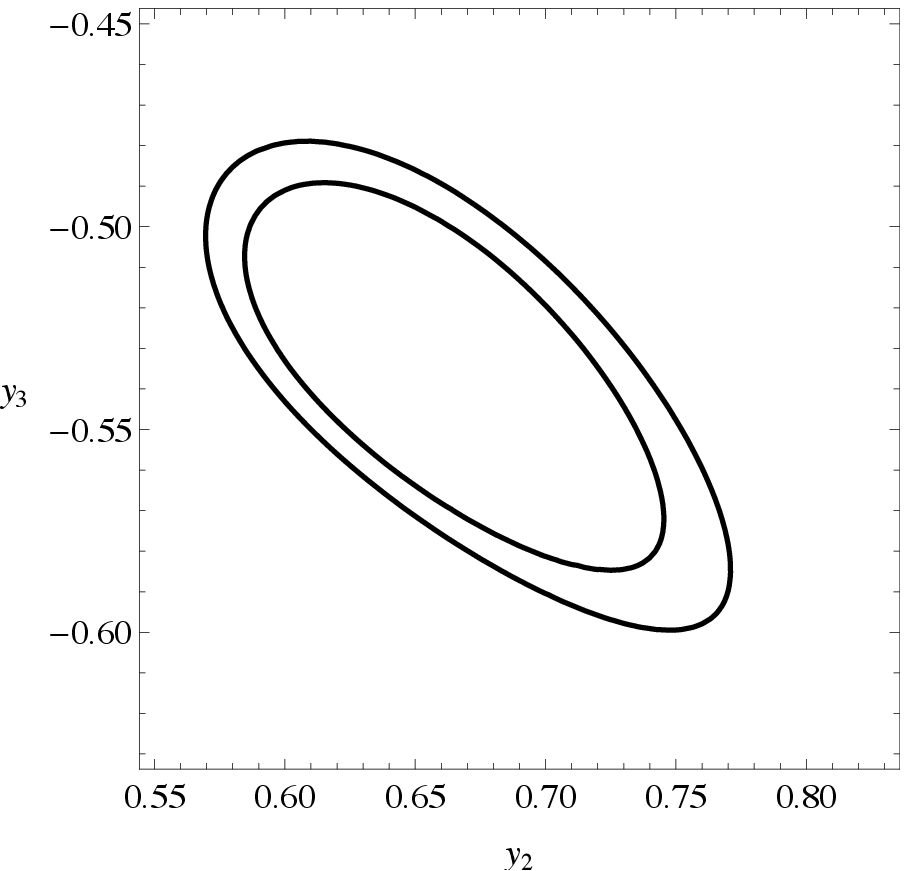}
\includegraphics[width=0.33\textwidth]{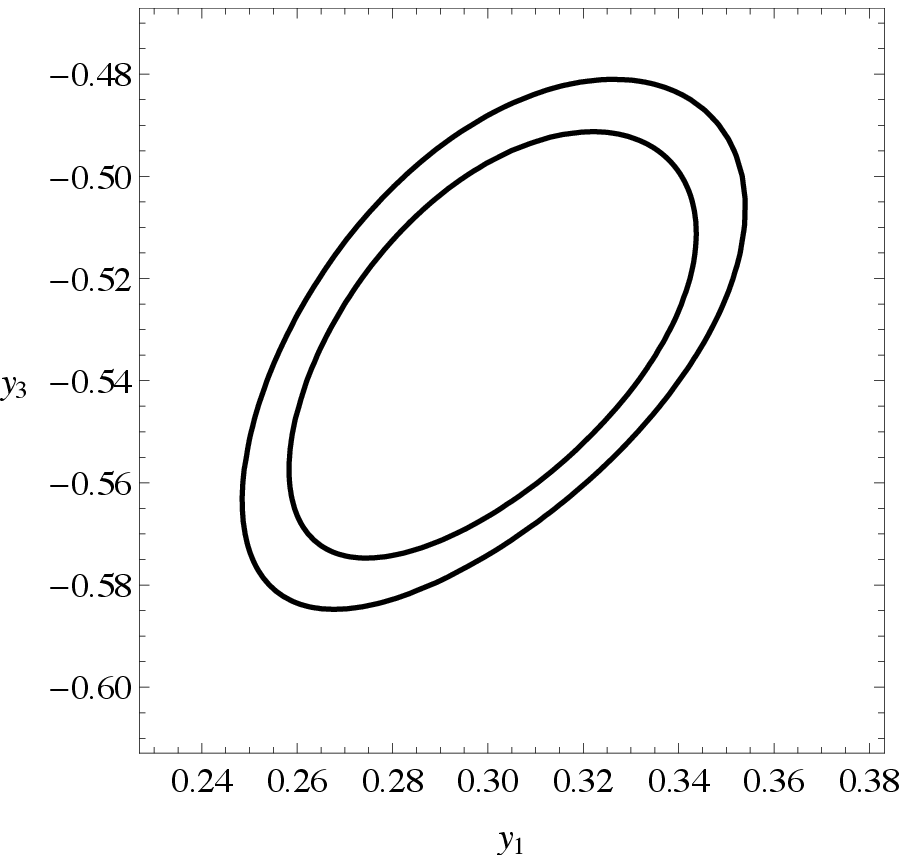}
\includegraphics[width=0.31\textwidth]{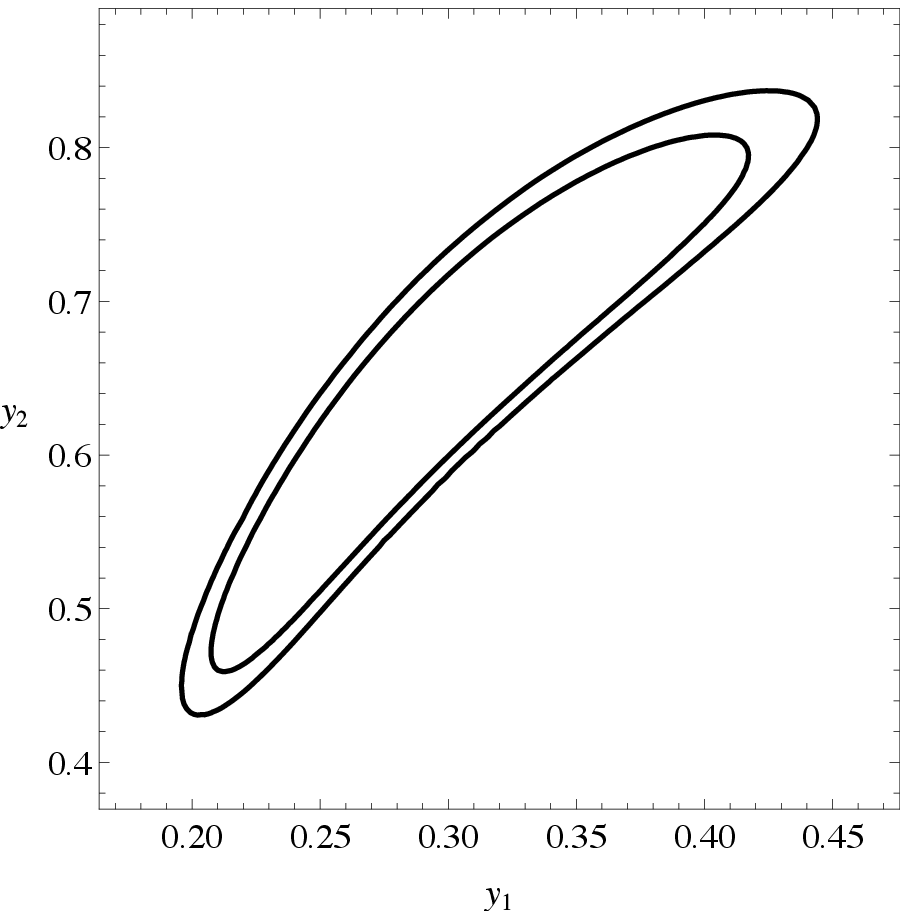}
\end{center}
\caption{\it 95\% CL (inside the inner ellipse) and 99\% CL (inside the outer ellipse) for $b_k=0.7$ in the plane $(y_2,y_3)$ with $y_1=y_3^0$ [left panel],  $(y_1,y_3)$ with $y_2=y_2^0$ [middle panel] and $(y_1,y_2)$ with $y_3=y_3^0$ [right panel].}
\label{b07}
\end{figure}

\begin{figure}[htb]
\begin{center}
\includegraphics[width=0.32\textwidth]{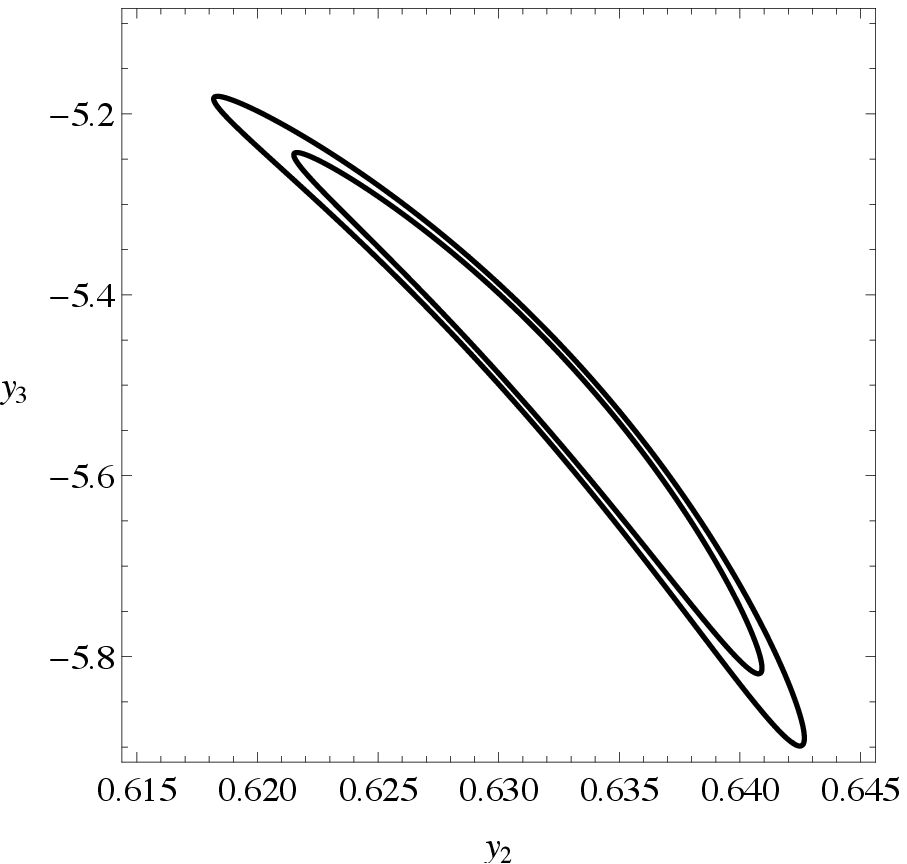}
\includegraphics[width=0.32\textwidth]{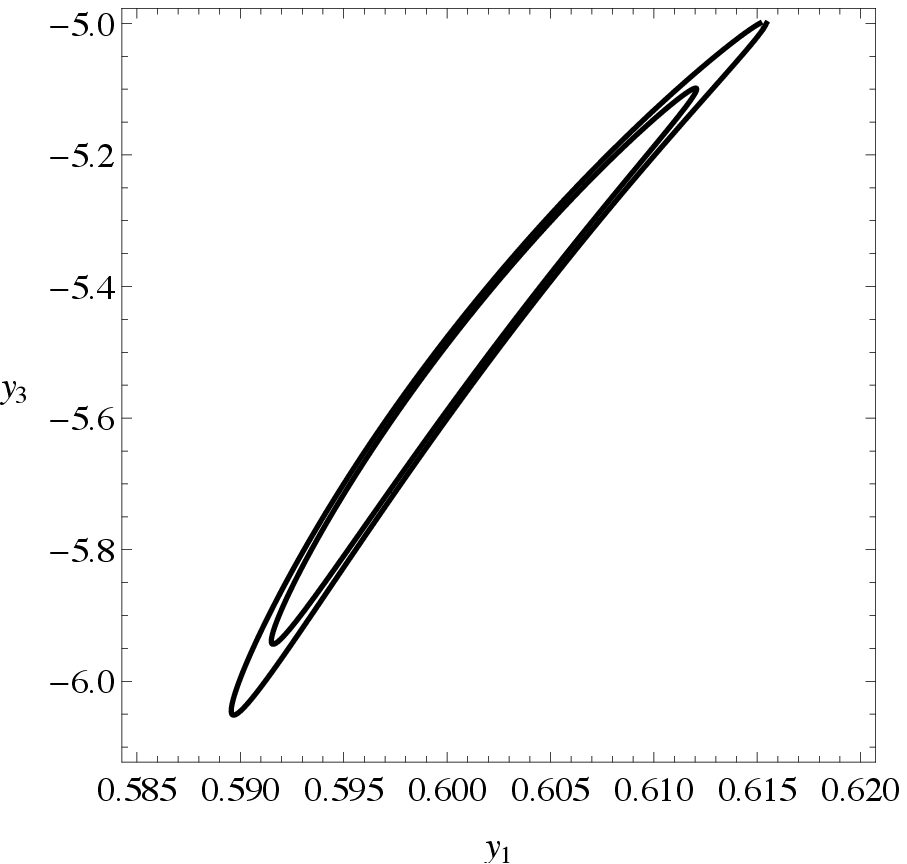}
\includegraphics[width=0.32\textwidth]{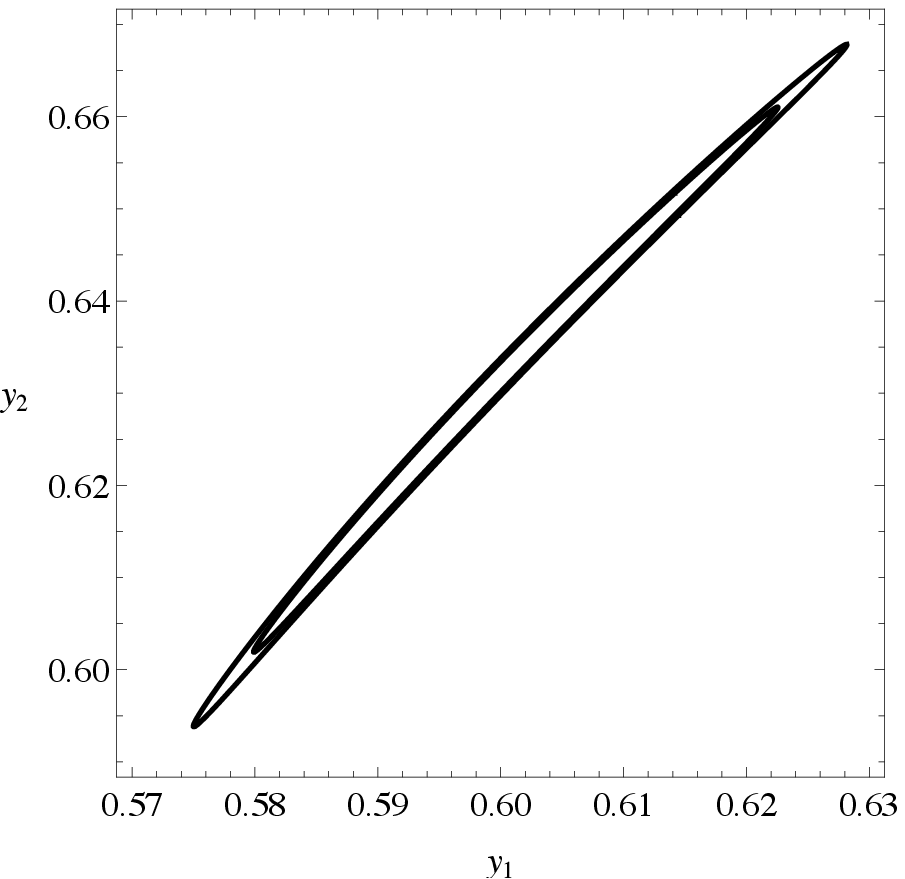}
\end{center}
\caption{\it 95\% CL (inside the inner ellipse) and 99\% CL (inside the outer ellipse) for $b_k=0.4$ in the plane $(y_2,y_3)$ with $y_1=y_3^0$ [left panel],  $(y_1,y_3)$ with $y_2=y_2^0$ [middle panel] and $(y_1,y_2)$ with $y_3=y_3^0$ [right panel].}
\label{b04}
\end{figure}

Another different case is provided by fixed values of $b_k$ in which case we can evaluate the available region in the space $y_k$ which is consistent with experimental data. 
We will now provide two simple examples yielding, respectively, a normal and inverted hierarchical neutrino spectrum. We first consider the case $b_k=0.7$ ($k=1,2,3$) such that $\chi^2_{min}\simeq 1.3$ for $y_k^0=(0.30,0.66,-0.53)$. In this case the neutrino spectrum at the best fit value is given by $m_i\simeq (0.004,0.010,0.050)$ eV which has a normal hierarchy but more hierarchical than the previous example. In Fig.~\ref{b07} the 95\% CL and 99\% CL regions are shown in the $(y_2,y_3)$ plane for $y_1=y_1^0$ (left panel), in the $(y_1,y_3)$ plane for $y_2=y_2^0$ (middle panel) and in the $(y_1,y_2)$ plane for $y_3=y_3^0$ (right panel). 
A second example yields an inverted hierarchical spectrum. In this case we are considering $b_k=0.4$ ($k=1,2,3$) such that $\chi^2_{min}\simeq 2$ for $y_k^0=(0.60,0.63,-5.5)$. In this case the neutrino spectrum at the best fit value is given by $m_i\simeq (0.057,0.058,0.031)$ eV which exhibits an inverted hierarchy. In Fig.~\ref{b04} the 95\% CL and 99\% CL regions are equally shown in the $(y_2,y_3)$ plane for $y_1=y_1^0$ (left panel), in the $(y_1,y_3)$ plane for $y_2=y_2^0$ (middle panel) and in the $(y_1,y_2)$ plane for $y_3=y_3^0$ (right panel). 

Finally let us comment on the size of the angles $b_i$ necessary to accommodate the observed neutrino data.
The values of the $b_i$ we used in the fit are somewhat small (but not excessively so)  compared to the full available parameter space $\sum_i b_i^2\leq \pi$, and one could be concerned about fine tuning. However we do not believe that there is a deep reason for that.  There are in fact many choices of parameters which yield a good fit, some of which at larger $b_i$.
However our approach of fitting does not allow for an honest estimation of the fine tuning, since the angles $b_i$ are dynamical variables. The only way to really assess the fine tuning is to compute the Coleman Weinberg potential, find the minimum as a function of the parameters (5D Yukawas and 5D masses), and compute the sensitivity with respect to these parameters. 

\section{Lepton Flavor Violation and Phenomenology}
\label{LFV}

Warped/composite models explaining  lepton masses and mixings  
are usually subject to severe constraints from lepton flavor violating processes. These can be mediated at tree level by the exchange of KK modes of electroweak gauge bosons ($\mu\to 3e$, $\mu-e$ conversion) as well as via loop diagrams ($\mu\to e \gamma$) that also involve the KK states of charged leptons, neutrinos and the Higgs boson. The simplest models, either with Dirac or Majorana neutrinos, are required to have KK scales in the 10 TeV region for $\mathcal O(1)$ Yukawa couplings~\cite{Kitano:2000wr,Huber:2003tu,Agashe:2006iy}~\footnote{We caution the reader that the quoted numbers have to be taken with a grain of salt. Scanning over the $\mathcal O(1)$ 5D Yukawa couplings typically results in a broad distribution of allowed KK scales, as has been shown for instance in the case of the quark flavor violation~\cite{Cabrer:2011qb}. Allowing for a moderate fine tuning can significantly reduce the bounds.}.
Moreover there is a tension between the tree-level and one-loop induced processes~\cite{Agashe:2006iy}: while tree level mediated FCNC's benefit from large 5D Yukawa couplings (allowing the zero modes to be more UV localized/elementary, and hence to decouple from the gauge KK modes), loop contributions to $\mu\to e \gamma$ naturally grow with the Yukawa coupling.
Various authors have thus tried to build models that reduce lepton flavor violation (LFV) via the introduction of either discrete~\cite{Csaki:2008qq} or continuous lepton flavor symmetries~\cite{Chen:2008qg}~\footnote{We also would like to mention the possibility to reduce the bounds in the context of soft-wall models~\cite{Atkins:2010cc}.}.
The model developed in Sec.~\ref{flavor} has a large symmetry and can hence naturally suppress LFV processes.
In fact the entire composite sector is completely symmetric under lepton family number since its bulk-to IR breaking is given by $U(3)_\ell\otimes U(3)_{\n}\otimes_i U(1)_{\e^i}\to \otimes_i U(1)_{\ell^i+\e^i+\n^i} $ and only the elementary sector (UV brane boundary conditions) breaks it. Since however the Higgs field is highly composite the elementary (UV brane localized) Yukawa interactions are completely negligible. As a consequence the charged lepton Yukawa couplings are simultaneously diagonal with the bulk masses for doublets ($c_\ell$), singlets $(c_\e)$ and RH neutrinos ($c_\n$). On the other hand the  UV BC for the RH neutrinos breaks this symmetry and introduces the Lepton mixing. This can be seen in various ways
\begin{itemize}
\item
The UV boundary conditions are aligned with the bulk masses and IR brane Yukawas ($n_0, Y_\n$ and $c_\n$ are all diagonal), and the breaking results from a nonzero VEV of $A_5$, the Hosotani breaking of the $U(3)_\n$ symmetry. 
\item
The VEV for $A_5$ is zero, $Y_\n$ and $c_\n$ are diagonal, but the Majorana brane mass (the UV boundary conditions for the RH neutrinos)  becomes non diagonal, $n_0^{-1}\to \widehat U^T\,n_0^{-1}\widehat U$.
\item
The VEV for $A_5$ is zero, $n_0$ and $c_\n$ are both diagonal, while $Y_\n$ becomes non-diagonal, $Y_\n\to \widehat UY_\n$.  
\end{itemize}
These interpretations are related by a 5D gauge transformation and are completely equivalent.

In all three cases, $c_\e$, $c_L$ and $Y_\e$ are diagonal. As a direct consequence, the mass and interaction eigenstates of the charged leptons are identical, {\em all} couplings to electroweak KK gauge bosons preserve flavor, and there are no tree-level mediated FCNC's.

\begin{figure}[htb]
\begin{center}
\SetScale{1.}
\begin{picture}(140,100)(0,0)
\SetWidth{1.}
\Line(0,50)(140,50)
\DashCArc(70,50)(40,0,180){3}
\Photon(70,50)(100,10){3}{5}
\Text(70,0)[t]{(a)}
\end{picture}
\hspace{1.5cm}
\begin{picture}(140,100)(0,40)
\SetWidth{1.}
\Line(0,50)(140,50)
\DashCArc(70,50)(40,0,180){3}
\Photon(70,90)(100,130){3}{5}
\Text(70,40)[t]{(b)}
\end{picture}
\end{center}
\caption{\it Diagrams contributing to $\mu\to e\gamma$. The solid line represents a fermion which is either a  charged lepton (a) or a neutrino (b), and the dashed line represents a boson which is either a neutral Higgs or $Z$ (a) or a charged (KK) Higgs or $W$ (b). }
\label{muegamma}
\end{figure}
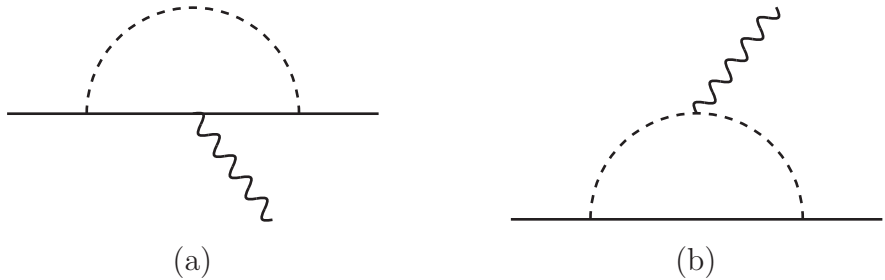

\begin{figure}[htb]
\begin{center}
\SetScale{1.}
\begin{picture}(140,100)(0,40)
\SetWidth{1.}
\Line(0,50)(140,50)
\PhotonArc(70,50)(40,0,180){3}{12}
\Photon(70,90)(100,130){3}{5}
\Vertex(15,50){2}
\Vertex(55,50){2}
\Vertex(85,50){2}
\Text(15,37)[]{$Y_\e$}
\Text(55,37)[]{$Y_\n$}
\Text(85,37)[]{$Y_\n$}
\Text(30,87)[]{$W^\pm$}
\end{picture}
\end{center}
\caption{\it The leading diagram contributing to $\mu\to e\gamma$ in the mass insertion approximation. }
\label{muegamma2}
\end{figure}
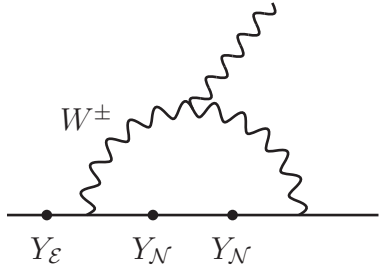

As far as the one-loop contributions to $\mu\to e\gamma $ are concerned, 
there are two types of diagrams, depicted in Fig.~\ref{muegamma}, depending on whether electric charge flows along a fermion line (a) or a boson line (b). Diagrams of type (a) never contribute in our model since all couplings and masses are simultaneously diagonal (no rotations on the charged fermions are ever necessary).
To understand the contribution from diagram (b) it is most convenient to work in the mass-insertion approximation and evaluate it in the basis of diagonal $n_0$ where all flavor violation is encoded in $Y_\n$ (third point of view above). Effectively we are expanding the diagram in powers of the 5D Yukawa coupling.
 The leading contribution from these diagrams have one or three Yukawa induced mass insertions  
 \cite{Csaki:2010aj}, however it is easy to convince oneself that the diagrams with one insertion always involve $Y_\e$. The leading diagram with three Yukawa insertions is depicted in Fig.~\ref{muegamma2} and 
  leads to the bound 
 \cite{Csaki:2010aj}
\be
a\,Y_*^2\left(\frac{3{\rm \ TeV}}{m_{\rm KK}}\right)^2\leq 0.015
\ee
where $Y_*$ is the typical size of the 5D Yukawa (in units of $k$) and $a\sim\mathcal O(5\%)$. 
Hence realistic KK scales of $\sim 2-3$ TeV can easily be accommodated with $Y_*\sim 0.4-0.5$. 
Notice that, due to the absence of tree level FCNC's there is no lower bound on $Y_*$.

Having shown the promising features of the proposed model in lepton flavor violating processes, we would like to close this section with 
some thoughts on the phenomenology of the additional states in the new gauge sector.
The lightest new vector resonances in our model arise from fields with Neumann BC in the UV and Dirichlet BC in the IR ($+-$ fields). Such fields have a light state of mass $m^{ND}_{KK}\approx 0.24\,ke^{-k\,y_1}$. In contrast, the lightest states of the $-+$ fields have $m^{DN}_{KK}\approx 2.4\,ke^{-k\,y_1}$, while those of $--$ fields have $m_{KK}^{DD}\approx 3.8\,k\,e^{-ky_1}$. 
On the other hand, since leptons are near UV localized, the coupling of the new vector resonances to $e$, $\mu$ and $\tau$ is dictated by their UV boundary conditions. Only KK modes of gauge fields with Neumann BC in the UV will have a significant coupling to leptons. For the light $+-$ state given above this coupling is approximately $ 0.17\, \hat g_\alpha$, where $\hat g_\alpha\equiv g_\alpha^{5D}\,k^{1/2}$ and $\alpha$ runs over the various factors of the bulk gauge group. For the electroweak gauge bosons the 5D gauge coupling is fixed in order to correctly reproduce the vector boson masses. However in the case of the lepton flavor gauge fields there is no such constraint and we can treat $g_\alpha^{5D}$ or equivalently $\hat g_\alpha$ as free parameters.

An obvious question is whether these fields could contribute significantly to electroweak precision observables at LEP. As the new gauge sector does not couple to the Higgs field we can safely neglect its contributions to the $S$ and $T$ parameters and focus on LEP2 cross section measurements at high energies (i.e.~above the mass poles of $W$ and $Z$). 
Since the present model does not have new gauge fields of $++$ type, and the $--$ and $-+$ fields have negligible couplings to leptons, the only fields of concern are of the $+-$ type, in particular the light mode mentioned above.  
Focusing on operators involving only electrons and positrons, the only $+-$ fields that contribute are $(A_\mu^{\ell,\alpha}-A_\mu^{\e,\alpha})/\sqrt{2}$ with $\alpha=0,3,8$. Using the exact expression for the sum over the $+-$ KK tower \cite{Cabrer:2011fb}, one obtains the following four-electron operators \footnote{For simplicity we take a single gauge coupling for $U(3)_\ell$ and $U(1)^3_\e$ respectively. Moreover, since $c_\ell,\ c_{\e^1}>1/2$ we can very well approximate the electrons by UV localized fields and one obtains $\sum_n [f^{+-}_n(0)/m_n^{+-}]^2 = e^{2ky_1}/2\,k^2$ which is almost entirely saturated by the lightest mode.}
\be
\mathcal L^{4e}=\frac{1}{16}(k e^{-ky_1})^{-2}  \left(
\hat g_\ell\, \bar e_L\gamma^\mu e_L-\hat g_\e\,\bar e_R\gamma^\mu e_R\right)^2
\label{4e}
\ee
Similar flavour-preserving operators are generated involving $\mu$ and $\tau$. 
Present bounds on lepton contact interactions can be found in Refs.~\cite{Nakamura:2010zzi,Bourilkov:2001pe}. These are based on an effective Lagrangian of the form
\be
\mathcal L=-\frac{2\pi}{\Lambda^2}\left(\eta_{LL}\,\bar e_L\gamma^\mu e_L\,\bar e_L\gamma_\mu e_L+\eta_{RR}\,\bar e_R\gamma^\mu e_R\,\bar e_R\gamma_\mu e_R+2 \eta_{LR}\,\bar e_L\gamma^\mu e_L\,\bar e_R\gamma_\mu e_R\right)
\ee
where $\eta_{\chi\chi'}=\pm 1,0$.  
Ref.~\cite{Bourilkov:2001pe} gives separate bounds for the cases
\be
\begin{array}{ccc}
\eta_{RR}=\eta_{LR}=0\,,\ \eta_{LL}=-1&&\Lambda>10.3 {\ \rm TeV}\\
\eta_{LL}=\eta_{LR}=0\,,\ \eta_{RR}=-1&&\Lambda>10.2 {\ \rm TeV}\\
\eta_{LL}=\eta_{RR}=-\eta_{LR}=-1&&\Lambda>16.5 {\ \rm TeV}
\end{array}
\ee
Fixing a typical IR scale $k\,e^{-ky_1}=1.25$ TeV (yielding 3 TeV SM resonances) we can then place bounds on the $\hat g_\alpha$
in several limiting cases
\be
\begin{array}{ccc}
\hat g_\e\ll \hat g_\ell&&\hat g_\ell<1.2 \\
\hat g_\ell\ll \hat g_\e&&\hat g_\e<1.2  \\
\hat g_\ell\approx \hat g_\e&&\hat g_\ell<0.76 
\end{array}
\ee
Although direct detection of the lightest $+-$ states at LHC seems impossible due to the absence of direct couplings to colored states, prospects at future linear colliders are much better. 
Indeed, in case LHC finds resonances of SM gauge bosons, it can be expected that the mass of the lightest $+-$ state be within the reach of the ILC with $\sqrt{s}=500$ GeV.

Another interesting question is the phenomenology of the scalars corresponding to the stabilized Wilson line moduli $A_5$. Such states have loop suppressed masses and can hence show up close to the electroweak scale~\footnote{The analogy of 5D composite Higgs models is evident. In fact the phenomenology resembles that of a pseudo Goldstone Higgs coupling only to leptons and heavy vector resonances.}. However, it is important to realize that in warped space they have negligible couplings to leptons (besides possibly the $\tau$), as they possess strongly IR localized profiles. Tree level production of such particles at a future linear collider should then proceed via fusion of the new gauge bosons (in complete analogy to vector boson fusion production of the Higgs boson in the SM), while decays to $\tau'$s will be strongly enhanced with respect to $\mu'$s and $e'$s. A detailed study of signatures of these new light degrees of freedom at lepton colliders is beyond the scope of the present paper and is left for future work. 

Finally, an interesting feature are loop effects such as contributions to the muon anomalous magnetic moment. 
These are generated via lepton flavor conserving penguin diagrams and have recently been computed for warped models in Ref.~\cite{Beneke:2012ie}. Given that the EW KK contributions are somewhere close to the observed deviation and that leptonic gauge couplings can be close to the weak ones, it would be interesting to perform a detailed study.

\section{Conclusion}
\label{conclusion}

Warped extra dimensional models provide an elegant theory of flavor by using the fact that a 5D Dirac mass localizes its fermion zero mode along the extra dimension in such a way that different localization for different fermions (i.e.~different 5D masses $c_f$) can account for the experimentally observed spectrum and mixing angles. Moreover FCNC higher dimensional operators involving light fermions (i.e.~fermions localized towards the UV brane, $c_f>1/2$) and generated by exchange of gauge boson KK-modes are protected by the so-called RS-GIM mechanism as massive KK-modes are leaning towards the IR brane. This mechanism which can account for hierarchical masses and mixing angles is a very appropriate one for describing the mass spectrum and mixing angles (described by the CKM matrix) in the quark sector if the constants $c_f$ are different in the left-handed quark doublet $c_{q_L}$ and right-handed up quark singlet $c_{u_R}$ sectors, while they are very degenerate in the right-handed down quark singlet $c_{d_R}$ sector, which suggests some protection by a flavor symmetry in the quark sector similar to that proposed in the present paper in the leptonic sector.

However in the lepton sector the situation is different as the charged lepton masses are hierarchical while neutrino masses and mixing angles (described by the PMNS matrix) follow an anarchic pattern. The localization mechanism does not work unless a fine-tuning on 5D masses is done, or it is implemented in a natural way by the symmetries of the theory. Moreover, unlike in the quark system, the nature of neutrinos, i.e.~Dirac versus Majorana, is not yet unveiled by experiments and both situations should be considered in model building. 

In the first part of the paper we have made a systematic review of the calculation of the neutrino mass matrix, both in the Dirac and Majorana cases, so that we can classify all possible cases which could give rise to realistic spectra and mixing angles. Similarly to what happens in 4D if lepton number is conserved in the bulk and in both branes neutrinos are Dirac fermions, while otherwise they are Majorana particles. For the case of Majorana neutrinos we have integrated out the 5D right-handed Majorana neutrinos in the process $\ell_i H\to \ell_j H$, a procedure similar to the seesaw mechanism in 4D.  In all cases the Yukawa matrix $Y_\n$ coupling the left-handed doublets $\ell_i$ with the right-handed neutrinos $\n_j$ can be a 5D (bulk) one and/or localized in either brane. Similarly lepton number can be violated in the bulk and/or in either brane. In both cases, Dirac or Majorana, the hierarchical charged lepton spectrum should be implemented by a  pattern in the corresponding 5D masses $c_\e^i$. Moreover a realistic spectrum and mixing angle pattern for neutrinos requires: 
\begin{itemize}
\item
For both Dirac and Majorana neutrinos: 
\begin{itemize}
\item
A bulk symmetry implementing that $c_\ell^i\equiv c_\ell$ and $c_\n^j\equiv c_\n$ independently on $i,j$.
\item
A Yukawa matrix $Y_\n$ with non-vanishing components along the bulk and/or the IR brane. A UV localized Yukawa matrix alone would provide too small neutrino masses.
\end{itemize}
\item
For Majorana neutrinos: Lepton number should not be violated in the bulk. Otherwise the charged lepton spectrum whould lead to a too heavy neutrino spectrum. Lepton number violating effects are thus dominated by those from the UV brane. 
\end{itemize}

In the second part of the paper we have constructed a simple model leading to the above required pattern for 5D masses, Yukawa couplings and lepton number violation. In short what we need is a bulk gauge group $G$ broken by boundary conditions to the subgroup $H_0$ ($H_1$) on the UV (IR) brane such that:
\begin{itemize}
\item
The space of zero modes for $A_5$, i.e.~the coset space $K=K_0\cap K_1$ $(K_i=G/H_i)$, is non-vanishing, to allow for non-trivial Wilson lines.
\item
The space of zero modes for $A_\mu$ is null, i.e. $H\equiv H\cap H_1=\varnothing$ to avoid unwanted massless non-SM gauge bosons.
\end{itemize}
For nontrivial $K$ one can still gauge away $A_5$ leading to misaligned BC for the RH neutrino at the two branes. The Majorana mass matrix will then depend on the WL and lead to nontrivial mixing. A priori the background $\langle A_5\rangle$ is a flat direction at tree-level (a classical modulus) which will however be dynamically determined at one-loop by the Coleman-Weinberg effective potential (the Hosotani mechanism). This will then result in a dynamical determination of the Majorana neutrino mass matrix. Computing the one-loop radiative corrections is to a large extent model dependent and it is outside the scope of the present paper. Here we try to stress the general features of the proposed mechanism while we postpone a study of a model dynamically implementing the neutrino mass matrix for further studies.

Without trying to classify all possible models we have proposed a model with $G=U(3)_\ell\otimes U(3)_\n\otimes_{i}U(1)_{\e^i}$, $H_1=\otimes_iU(1)_{(\ell+ \e+ \n)^i}$ and $H_0=U(3)_\ell\otimes U(1)_{\lambda^1_\n}\otimes_i U(1)_{\e^i}$ which leads to lepton number violation on the UV brane (with a particular pattern) and diagonal IR localized Yukawa matrices $Y_\n$ and $Y_\e$, although we point out that other choices could do a similar job. In this particular example the non-trivial coset  contains $SO(3)_\n\subset SU(3)_\n$  and, for diagonal UV localized Majorana mass matrix and unit Yukawa matrix $Y_\n$, it can be identified with the three angles of the PMNS rotation. 
Fitting also the neutrino mass spectrum requires a diagonal (non-unit) matrix $Y_\n$ and the Wilson line depart from the PMNS angles. We have quantified the available region for the Wilson line parameters and found that for a 95\% CL region in the fit there is no fine-tuning.

In general the larger the bulk gauge group $G$ the more protection against FCNC's. We have analyzed in the present model lepton flavor violation induced by tree-level exchange of KK-modes (as $\mu\to 3e$ and $\mu\to e$ conversion) and one-loop (as $\mu\to e\gamma$) processes. As a consequence of the fact that $c_\e$, $c_\ell$ and $Y_\e$ are diagonal the mass and interaction eigenstates for charged leptons coincide, all couplings to KK gauge bosons are flavor diagonal and there are no tree-level lepton flavor violating processes.  On the other hand lowering the rate of the process $\mu\to e\gamma$ below the experimental bound requires an upper bound on the typical value of the entries of the matrix $Y_\n$. In the absence of tree-level lepton flavor violating processes the usual tension disappears and it is possible to find realistic values for the loop-mediated processes for $\mathcal O(1)$ Yukawas and KK-masses about a few TeV.

Interestingly enough the common value for the left-handed lepton doublet constant $c_\ell \simeq~0.63$ agrees (within 1$\sigma$) with the common value of the right-handed down singlets $c_{d_R}\simeq 0.65$~\cite{Cabrer:2011qb} which might suggest an extension of the flavor symmetry to the quark sector commuting with some unification group. However the difficulty in extending the idea of breaking flavor symmetries and
creating fermion mixing by Wilson lines can be realized as follows. Quark mixing can only arise from the Yukawa structure, and the UV brane
Yukawa couplings are highly suppressed due to the large suppression of the Higgs wave function there. Hence, if the flavor symmetry is broken
only nonlocally, it will be impossible to generate a sizable Cabbibo angle. For the same reason the present mechanism is not applicable to generate Dirac neutrinos.
The only way out would be a sufficiently large breaking of the bulk flavor symmetries on the IR brane which would however lead to completely different type of models (for models where bulk and brane flavor symmetries are used in the quark sector see~\cite{Cacciapaglia:2007fw,Fitzpatrick:2007sa}).

Finally we would like to point out that we have performed all calculations in the present paper for an RS (AdS) 5D metric. However a similar study would also apply for IR deformed metrics (as the soft-wall class of metrics~\cite{Archer:2011bk,Cabrer:2009we}) and we expect similar conclusions to follow. In particular for brane localized Majorana masses and Yukawa couplings, as in the example we have worked out, most of the results (except for normalization factors) only depend on the total warp factor, which is metric independent while gauge breaking arguments should hold in general metrics.

\section*{Acknowledgment}

We thank C.~Biggio for participation in the earliest stages of this work. Work supported in part by the Spanish Consolider-Ingenio 2010 Programme CPAN (CSD2007-00042) and by CICYT-FEDER-FPA2011-25948. The research of GG is supported by the ERC Advanced Grant 226371, the ITN programme PITN-GA-2009-237920 and the IFCPAR CEFIPRA programme 4104-2.

\vspace{1cm}
\section*{Appendix}
\appendix

\section{Calculating the propagator $G_{RR}$ for $c_M\ne 0$}
\label{details}
In this section we give the derivation of expression (\ref{equ_GRR}). 
Using the equation of motion (\ref{Dirac2}) and the definition of $G_{RR}$ 
\be
G_{RR}=\sum_n\frac{\hat f_R^{(n)}(y)\,\hat f_R^{(n)}(y')}{m_n}\,,
\ee
 we can write
\begin{multline}
(M_D-\partial_y)M_M^{-1}(M_D+\partial_y)G_{RR}(y,y')\\
=(M_D-\partial_y)M_M^{-1}e^A\sum_n\hat f_L^{(n)}(y)\hat f^{(n)}_R(y')
+e^A \sum_n \hat f_R^{(n)}(y)\hat f^{(n)}_R(y')-M_M\, G_{RR}(y,y')
\end{multline}
Now note that the orthogonality and completeness relations are 
\bea
\int dy\, e^A\left[\hat f^{(m)}_L\hat f_L^{(n)}+\hat f^{(m)}_R\hat f^{(n)}_R\right]&=&\delta_{mn}\\
e^A\sum_n \left[\hat f^{(n)}_\chi (y)\hat f^{(n)}_{\chi'}(y')\right]&=&\delta_{\chi\chi'}\delta(y-y')
\label{completeness}
\,.
\eea
Assuming that there are no zero modes one can use the completeness relations to write
\be
\left[(M_D-\partial_y)M_M^{-1}(M_D+\partial_y)+M_M\right]G_{RR}(y,y')=
\delta(y-y')
\label{eqnprop}
\ee
Now we will make the simplifying assumption $M_M(y)=c_MM(y)$, $M_D(y)=c_\n M(y)$, where $M(y)$ is an arbitrary function. In other words, the 5D Dirac and Majorana masses have the same $y$ dependence. Then the general solution to Eq.~(\ref{eqnprop}) is
\be
G_{RR}^{\lessgtr}=\alpha^{\lessgtr} e^{\ceff Q}+\beta^{\lessgtr}e^{-\ceff Q}
\ee
where $G^<$ ($G^>$) refers to the regimes $y<y'$ ($y>y'$), $Q(y)=\int_0^y M$ and $\ceff=\sqrt{c_\n^2+c_M^2}$ .
Continuity of $G_{RR}$ at $y'$ gives 
\be
(\alpha_<-\alpha_>)e^{\ceff Q(y')}+(\beta_<-\beta_>)e^{-\ceff Q(y')}=0
\ee
The jump condition for $G'$ gives
\be
(\alpha_<-\alpha_>)e^{\ceff Q(y')}-(\beta_<-\beta_>)e^{-\ceff Q(y')}=\frac{c_M}{\ceff}
\ee
This can easily be solved to yield
\bea
G^<_{RR}(y,y')&=&b\, \frac{c_M}{\ceff}e^{\ceff[Q(y')-Q(y)]}+c\frac{c_M}{\ceff}e^{\ceff[Q(y)-Q(y')]}\,\label{equ_GRRsmallerlarger}\\
G^>_{RR}(y,y')&=&\left(b+\frac{1}{2}\right)\, \frac{c_M}{\ceff}e^{\ceff[Q(y')-Q(y)]}+\left(c-\frac{1}{2}\right)\frac{c_M}{\ceff}e^{\ceff[Q(y)-Q(y')]}\nn
\eea
The constants $b$ and $c$ are determined from the BC's at $0$ and $y_1$. Note they are functions of $y'$.
One can express the solutions in terms of
\be 
Q_m=\int\limits_0^{\min(y,y')} M\,,\qquad  Q_M=\int\limits_{\max(y,y')}^{y_1}M\,,\qquad Q_1=\int\limits_0^{y_1}M\,.
\ee
Let us consider the most general BC's, which we can write as
\bea
(c_0 M + \partial_y)G^{<}_{RR}(0)&=&0 \label{equ_generalBC}\\ 
(c_1 M + \partial_y)G^{>}_{RR}(y_1)&=&0\nn
\eea
In the limit $c_i\rightarrow \infty$ we recover the boundary condition where $\hat f_R(y_i)=0$ (here $y_0=0$). Taking $c_i=c_D$ however corresponds to $\hat f_L(y_i)=0$ or in other words $(M_D+\partial_y)\hat f_R=0$~\footnote{Some of these special cases have been investigated earlier \cite{Watanabe:2010cy}.}. 
Applying the BC's to (\ref{equ_GRRsmallerlarger}) we find the integration constants
\bea
b&=&\frac{1}{2}\frac{c_0+\ceff}{N}\left[(c_1-\ceff)e^{-\ceff Q_1} - (c_1+\ceff)e^{\ceff(Q_1-2Q(y'))}\right]\\
c&=&\frac{1}{2}\frac{c_0-\ceff}{N}\left[(c_1+\ceff)e^{\ceff Q_1} - (c_1-\ceff)e^{\ceff(-Q_1+2Q(y'))}\right]\nonumber\\
N&=&2\left[(c_0 c_1 - \ceff^2)\sinh(\ceff Q_1) + \ceff(c_0 - c_1)\cosh(\ceff Q_1)\right]\nonumber
\eea
and with these we get the final expression (\ref{equ_GRR})
\be
G_{RR}(y,y')= \frac{c_M}{\ceff} \frac{
 [c_0 \sinh(\ceff Q_m)-\ceff\cosh(\ceff Q_m)  ][c_1 \sinh(\ceff Q_M)+ \ceff\cosh(\ceff Q_M)]} {  (c_0 c_1 - \ceff^2)\sinh(\ceff Q_1) + \ceff(c_0-c_1)\cosh(\ceff Q_1)}
\ee
To give a meaning to the variables $c_i$ we use the equation of motion (\ref{Dirac2}) to eliminate $\hat f_L$ in the BC's (\ref{BC1}). This results in
\be
\left( m_n e^A n_i + (c_\n+c_M n_i)M + \partial_y \right) \hat f_R(y_i) = 0
\ee
Applying this to the definition of the propagator (\ref{def}) and using the completeness relation (\ref{completeness}) we get
\be
\left[(c_M n_i + c_\n) M + \partial_y\right]G_{RR}(y_i)=0
\ee
Comparing this to (\ref{equ_generalBC}) we see that we can identify $c_i$ as a function of the numbers $n_i$ of the boundary mass terms in (\ref{LB1}).
\be
c_i = c_M n_i + c_\n
\ee
\section{Calculation of $G_{RR}$ for $n$ RH neutrinos and $c_M=0$}
\label{moredetails}

In the case $c_M=0$ and $n$ right-handed neutrinos we get matrix equations. We will work in the basis of diagonal $c_\n$. The EOM (\ref{Dirac2}) has one term less than in the general case, due to the missing $c_M$ and indices are added to emphasize the matrix structure of $c_{\n}^{ab} = \delta^{ab} c_{\n}^b$ and $n_i^{ab}$. 
\be
m_n e^A \hat f^{(n)\,a}_{L/R}=(c^{a}_{\n}M \pm \partial_y)\hat f^{(n)\,a}_{R/L} \label{equ_EOMcMZeroNRHneutrinos}
\ee 
The BC's are given by
\be
\hat f^{(n)\,a}_{L}(y_i) + n_i^{ab}\hat f^{(n)\,b}_{R} = 0 \label{equ_BCcMZeroNRHneutrinos}
\ee
%

Multiplying the EOM (\ref{equ_EOMcMZeroNRHneutrinos}) by $\frac{\hat f^{(n)\,c}_{\chi}(y')}{m_n}$and taking the sum over $n$ we find 
\be
    e^A \sum_n \hat f^{(n)\,a}_{L/R}(y)\hat f^{(n)\,c}_{\chi}(y')
      =(c^{a}_{\n}M \pm \partial_y)\sum_n \frac{\hat f^{(n)\,a}_{R/L}(y) \hat f^{(n)\,c}_{\chi}(y')}{m_n}
 \ee
and using again the completeness relation  one gets
\be
 \delta_{L/R,\chi}\delta(y-y')\delta^{ac} = (c^{a}_{\n}M \pm \partial_y)G_{R/L,\chi}^{ac}(y,y')\label{equ_EOM_GRR}
\ee
where we have used the definition
\be
G_{R/L,\chi}^{ab}(y,y') = \sum_n\frac{\hat f^{(n)\,a}_{R/L}(y)\hat f^{(n)\,b}_\chi(y')}{m_n}
\ee
Solving this first order differential equation yields for $G_{RR}$
\be
G^{ab}_{RR}(y,y') = e^{-c^a_{\n}Q(y)} \alpha^{ab}(y') 
\ee
With $Q(y) = \int_0^y dz M(z)$. In contrast to the case of $c_M\ne0$ we do not have to worry about jump conditions as there is no delta funcion in the equation for $G_{RR}$.
Multiplying the BC's by $\hat f_R(y')^{(n)\,c}$ and taking the sum over $n$ results in a coupled equation for $G_{LR}$ and $G_{RR}$ on the boundary
\be
G_{LR}(y_i,y') + n_i G_{RR}(y_i,y') = 0 \label{equ_BC_GRR}
\ee
So in order to find $\alpha$  we have to find $G_{LR}$ first. From  the EOM (\ref{equ_EOM_GRR}) we find a differential equation with a delta function. Therefore we have to solve it for the two cases $y<y'$ and $y>y'$ and connect the resulting function with the jump condition, found by integrating the EOM around $y'$.
We can write
\be
G_{LR}(y,y') = G^<_{LR}(y,y') \Theta(y'-y) + G^>_{LR}(y,y') \Theta(y-y'), 
\ee
with the stepfunction 
\be
\Theta(y)=\left\{\begin{array}{c}1 \text{  for  } y>0\\0 \text{  for  } y<0\end{array} \right.
\ee
For the two cases $y\lessgtr y'$ we can derive
\be
G^\lessgtr_{LR} = e^{c_{\n} Q(y)} \beta^\lessgtr(y').
\ee
From the jump condition 
\be
{G^{ab}_{LR}}^>(y',y') - {G^{ab}_{LR}}^<(y',y') = \delta^{ab}
\ee
the relation for the constants $\beta^\lessgtr$ is found:
\be
\beta^< = e^{-c_\n Q(y')} + \beta^>.
\ee
With these expressions for $G_{LR}$ and $G_{RR}$ the BC's become
\be
e^{+c_\n Q(y_i)} \left( e^{-c_\n Q(y') +\beta^>}\Theta(y'-y) + \beta^>\Theta(y-y')  \right) + n_i e^{-c_\n Q(y_i)} \alpha = 0
\ee
which can be solved on the two branes with $y_0=0$ and result in
\be
\alpha = \left( e^{-c_\n Q(y_1)} n_1 e^{-c_\n Q(y_1)} - n_0 \right) e^{-c_\n Q(y')}.
\ee
Thus we find
\be
G_{RR}(y,y') = e^{-c_\n Q(y)} \left( e^{-c_\n Q(y_1)} n_1 e^{-c_\n Q(y_1)} - n_0 \right) e^{-c_\n Q(y')},
\ee
which in the case $M(y) = ky$ results in Eq.~(\ref{equ_GRR_cMzero}) in section~\ref{cMzero}.


%
%
%
%
%
%

\end{document}